\documentclass[11pt,a4paper]{article}
\usepackage[dvipdfmx]{graphicx}
\usepackage{jheppub}
\pdfoutput=1

\usepackage{amssymb,amsmath}

\usepackage{color}
\usepackage{float}
\usepackage{bbold}
\usepackage{enumitem}
\usepackage{multirow}
\usepackage{array}

\usepackage{hyperref}
\usepackage{cleveref}

\def\beq{\begin{equation}}
\def\eeq{\end{equation}}

\def\bea{\begin{eqnarray}}
\def\eea{\end{eqnarray}}

\def\bei{\begin{itemize}}
\def\eei{\end{itemize}}

\newcommand{\mx}{m_X}
\newcommand{\my}{m_Y}
\newcommand{\mz}{m_Z}

\newcommand{\mxs}{m_X^2}
\newcommand{\mys}{m_Y^2}
\newcommand{\mzs}{m_Z^2}
\newcommand{\mfs}{m_5^2}
\newcommand{\mf}{m_5}
\newcommand{\mzmf}{\frac{m_Z}{m_5}}
\newcommand{\mymz}{\frac{m_Y}{m_Z}}
\newcommand{\mymf}{\frac{m_Y}{m_5}}
\newcommand{\mxmy}{\frac{m_X}{m_Y}}
\newcommand{\mxmz}{\frac{m_X}{m_Z}}

\newcommand{\Dfour}{\Delta_4}

\newcommand{\LSB}{\left[}
\newcommand{\RSB}{\right]}

\abstract{
Collider events with multi-stage cascade decays fill out the kinematically allowed region in phase space with a density that is enhanced at the boundary. The boundary encodes all available information about the spectrum and is well populated even with moderate signal statistics due to this enhancement. In previous work, the improvement in the precision of mass measurements for cascade decays with three visible and one invisible particles was demonstrated when the full boundary information is used instead of endpoints of one-dimensional projections. We extend these results to cascade decays with four visible and one invisible particles. We also comment on how the topology of the cascade decay can be determined from the differential distribution of events in these scenarios.
}
\subheader{UTTG-20-16}
\title{Multi-Dimensional Phase Space Methods for Mass Measurements and Decay Topology Determination}

\author[a]{Baris Altunkaynak,}
\affiliation[a]{Department of Physics, Northeastern University, Boston, MA 02115, USA}
\emailAdd{i.altunkaynak@northeastern.edu}
\author[b]{Can Kilic}
\emailAdd{kilic@physics.utexas.edu}
\affiliation[b]{Theory Group, Department of Physics and Texas Cosmology Center,
\\The University of Texas at Austin,  Austin, TX 78712 U.S.A.}
\author[b]{and Matthew D. Klimek}
\emailAdd{klimek@physics.utexas.edu}

\notoc
\begin{document}
\maketitle
\flushbottom

\section{Introduction}
\label{sec:intro}

Naturalness of the Higgs sector as well as the weakly interacting massive particle (WIMP) paradigm for dark matter provide strong motivations for new physics at the TeV scale. The most commonly studied extensions of the Standard Model (SM) that attempt to solve the hierarchy problem do so by positing the existence of partners to the SM particles that cancel divergent contributions to the Higgs mass. Many of these scenarios also provide a dark matter candidate since they incorporate a parity symmetry under which the partner particles are odd, making the lightest partner particle stable. Arguably the best known example for such scenarios is the minimal supersymmetric extension of the SM, the MSSM.

The collider phenomenology of these scenarios has been studied extensively in the literature. The most promising discovery channels include production of colored partners, which then decay, often in multiple stages, until the lightest partner is reached. Since the lightest partner is assumed to constitute dark matter, it leaves the detector without interacting. Thus no resonances can be constructed from the visible decay products and discovery as well as mass measurement prospects often rely on endpoints of one-dimensional distributions of Lorentz invariant (e.g. edges, endpoints)
\cite{Hinchliffe:1996iu,Bachacou:1999zb,Allanach:2000kt,Gjelsten:2004ki,Gjelsten:2005aw,Lester:2005je,Miller:2005zp,Lester:2006yw,Ross:2007rm,Barr:2008ba,Barr:2008hv,Klimek:2016axq} (for a comprehensive review, see \cite{Barr:2010zj})
 or boost invariant (e.g. $m_{T}$, $m_{T2}$) variables \cite{
Lester:1999tx,
Barr:2003rg,
Meade:2006dw,
Baumgart:2006pa,
Matsumoto:2006ws,
Lester:2007fq,
Cho:2007qv,
Gripaios:2007is,
Nojiri:2008hy,
Tovey:2008ui,
Cho:2008cu,
Serna:2008zk,
Barr:2008ba,
Nojiri:2008vq,
Cho:2008tj,
Cheng:2008hk,
Choi:2009hn,
Matchev:2009ad,
Cohen:2010wv,
Polesello:2009rn,
Konar:2009wn,
Konar:2009qr,
Curtin:2011ng,
Nojiri:2010mk,
Lester:2011nj,
Mahbubani:2012kx}.

The Large Hadron Collider (LHC) has completed its 7 and 8~TeV runs and is currently running with a center of mass energy of 13~TeV. The LHC experiments currently do not have significant indications of physics beyond the SM. Considering that the center of mass energy is already near the design value, one needs to take seriously the possibility that if new physics is discovered by the LHC experiments, the signal statistics will remain low, or moderate at best. Therefore, it will be of paramount importance to optimize the methods by which the signal will be studied for low statistics. 

Let us consider mass measurement techniques in particular. For cascade decay chains with sufficiently many intermediate on-shell stages, polynomial methods~
\cite{
Hinchliffe:1998ys,
Nojiri:2003tu,
Kawagoe:2004rz,
Gjelsten:2006as,
Cheng:2007xv,
Nojiri:2008ir,
Cheng:2008mg,
Cheng:2009fw,
Matchev:2009iw,
Autermann:2009js,
Kim:2009si,
Han:2009ss,
Kang:2009sk,
Webber:2009vm,
Nojiri:2010dk,
Kang:2010hb,
Hubisz:2010ur,
Cheng:2010yy,
Gripaios:2011jm,
Han:2012nm,
Han:2012nr}
 can be applied to algebraically solve for all unknown masses based on a small number of events. However, there exist decay chains which do not have sufficiently many on-shell stages for these methods to be applicable. For such decay chains, the one-dimensional variables mentioned above are commonly accepted as the tool to be used for mass measurements. It was argued in ref. \cite{Agrawal:2013uka} however that when there are more than two visible particles in the final state, the kinematically accessible region in phase space is multidimensional and the commonly used one-dimensional variables are inefficient at low statistics. It was demonstrated specifically for final states with three visible particles and one invisible particle that the density of events near the boundary of the kinematically accessible phase space is enhanced, and that a determination of this boundary in the multidimensional phase space could yield significantly higher precision and accuracy for mass measurements at low statistics.

In this paper we will extend the conclusions of ref. \cite{Agrawal:2013uka} to the remaining cascade decay topologies where polynomial methods are not applicable. If all on-shell decay stages are 2 or 3-body decays with one invisible particle emitted from the last stage of the cascade, then it is straightforward to show that any cascade decay with more than five final state particles can be analyzed using polynomial methods, therefore we will restrict ourselves to final states with at most five final state particles. We will show that the enhancement in the density of events near the boundary is in fact even stronger for five-body decays compared to four-body decays, and in a number of representative cases for decay topologies we will demonstrate the improvement for mass measurements compared to the more traditional methods based on kinematic edges or endpoints. 

Various techniques involving kinematic variables have also been proposed for the purpose of determining decay topologies \cite{Agashe:2010gt,Agashe:2010tu,Agashe:2012fs,Cho:2012er,Cho:2014naa,Dev:2015kca,Kim:2015bnd}.
We will provide a preliminary assessment of the sensitivity of the full phase space boundary method to the topology, and suggest an algorithm by which the topology underlying a signal sample should be determined.

Our goal will be to provide a proof of principle that these improvements can be obtained, and therefore as in ref. \cite{Agrawal:2013uka} we will compare our methods to those based on kinematic endpoints under ideal circumstances, without SM or combinatorial backgrounds, spin correlations or realistic detector effects. While these certainly pose additional challenges in the construction of a fully realistic analysis, they will deteriorate the results of both our methods and any method based on kinematic endpoints, with no obvious reason why one should be more negatively affected than the other. Also, as in \cite{Agrawal:2013uka} we will restrict our study to ``one-sided'' events, where the cascade decay takes place on one side of the event, and the other side is assumed to include only the lightest partner. This corresponds to scenarios such as gluino-LSP associated production in the MSSM. The reason for this choice is that our methods use only Lorentz-invariant observables and are therefore used on one decay chain at a time, with no obvious way to combine the two sides of the event using the missing transverse energy (MET) for example. Therefore, for reasons of simplicity, we demonstrate the applicability of our methods in the simplest possible case of one-sided events. The same methods can of course simply be used {\it twice} in a symmetric event, but that comes at the cost of combinatoric issues such as identifying which side of the event any final state particle belongs to. We will leave a more realistic study including all these complications to future work. In fact, in parallel to this work, methods are already being developed to address some of these complications, and for one decay topology featured in ref. \cite{Agrawal:2013uka} it has already been demonstrated \cite{Debnath:2015wra,Debnath:2016mwb,Debnath:2016gwz} that the improvement for mass measurements based on the determination of the full phase space boundary over one-dimensional variables can be maintained in the presence of SM and combinatorial backgrounds, by using Voronoi tessellations.

The layout of the paper is as follows. In section~\ref{sec:phasespace} we review the mathematical description of many-body phase space and we quantify the enhancement near the boundary for five-body final states. In section~\ref{sec:mass} we focus on mass measurements and we set up an analysis to compare the results of mass measurement based on our methods to those obtained from kinematic endpoints. In section~\ref{sec:topology} we comment on the potential use of our methods for determining the underlying decay topology. We conclude in section~\ref{sec:conclusions}. Certain details of our methods are more fully described in appendices \ref{app:endpoints} through \ref{app:factorization}.


\section{Mathematical Description of Many-Body Phase Space}
\label{sec:phasespace}

The standard form of the phase space volume element of $n$ final state particles with 4-momenta $p_{i}^{\mu}$ and total 4-momentum $P_{\mu}$
\begin{eqnarray}
	dPS_n &=&  \left(\prod_{i=1}^n \frac{d^4 p_i}{(2\pi)^3} \delta(p_i^2 - m_i^2) \right) (2\pi)^4 \delta^4\left(\sum_{i=1}^n p_i^\mu - P^\mu\right) \nonumber \\
	&=&  \left(\prod_{i=1}^n \frac{d^3 \mathbf p_i}{(2\pi)^3 2E_i} \right) (2\pi)^4 \delta^4\left(\sum_{i=1}^n p_i^\mu - P^\mu\right) 
\end{eqnarray}
is expressed as a function of individual components of 4-momenta which are not manifestly Lorentz invariant. There also exists a less well-known formulation which is expressed purely in terms of Lorentz scalars~\cite{Byckling:1971vca,RevModPhys.36.595}.  As argued in~\cite{Agrawal:2013uka} this form contains important clues to optimizing the sensitivity of mass measurements, therefore we will review it below.

We start by defining $M_n$ as the $n\times n$ matrix with elements $p_{i}\cdot p_{j}$, and define $\Delta_{i}$ as the coefficients of the characteristic polynomial of $M_{n}$ as follows:
\begin{equation}
{\rm Det}\left[\lambda 1_{n\times n}-M_{n}\right]\equiv \lambda^{n}-\left(\sum_{i=1}^{n}\Delta_{i}\lambda^{n-i}\right).
\end{equation}
The kinematically accessible region of phase space corresponds to $\Delta_{1,2,3}>0$, $\Delta_{4}\ge 0$ and $\Delta_{5,\ldots ,n}=0$, with $\Delta_{4}=0$ defining the boundary of this region~\cite{RevModPhys.36.595}. For the specific case of $n=4$, the volume element is given by
\beq
	dPS_4 = ({\rm const.})\times M_X^{-2} \left(\prod_{i<j}dm^2_{ij}\right) \Dfour^{-1/2}\Theta(\Dfour) \delta\left(\sum_{i<j} m^2_{ij} - {\rm const.}\right) ,
\eeq
where $M_{X}^{2}=P_{\mu}P^{\mu}$ and where the $\delta$-function at the end enforces energy conservation. Note that the volume element scales as $\Dfour^{-1/2}$, diverging near the boundary in an integrable way. This can be understood as follows: $\Dfour$, which for $n=4$ is equal to $(-\det M_4)$ can be rewritten as $-\det(V^T g V) = \det^2 V$, where $V$ is the $4\times 4$ matrix whose columns are the $p_{i}^{\mu}$ and $g=\text{diag}(1,-1,-1,-1)$ is the metric. This makes it clear that the boundary of the kinematically accessible region corresponds to the final state momenta becoming linearly dependent.  When this happens, the coordinate change from Cartesian coordinates to the Lorentz-invariant coordinates $m_{ij}^{2}$ becomes singular and the Jacobian diverges. Note also that the presence of intermediate on-shell particles in the cascade does not change this conclusion, since in the narrow width approximation, these contribute $\delta$-functions to the amplitude squared $|{\mathcal M}|^{2}$, the arguments of which are linear in the $m_{ij}^{2}$. Therefore, using these $\delta$-functions to eliminate some of the integrals over $m_{ij}^{2}$ never produces nontrivial Jacobian factors.

Going beyond $n=4$, the phase space volume element has the form~\cite{RevModPhys.36.595}
\beq \label{yb_ps_delta}
	dPS_n = ({\rm const.}) \times M_X^{-2}\,\left(\prod_{i<j}dm^2_{ij}\right) \Dfour^{(n-5)/2}\Theta(\Dfour)\delta(\Delta_5)\cdots\delta(\Delta_n)  \delta\left(\sum_{i<j} m_{ij}^{2} - {\rm const.}\right).
\eeq
Naively, this expression seems to imply that the enhancement in the volume element near the boundary is absent for $n > 4$. 
However, a more careful examination reveals that the arguments of the $\delta(\Delta_n)$ factors are non-linear in the $m_{ij}^2$, and therefore nontrivial Jacobians arise as those $\delta$-functions are integrated over.  

\begin{figure}
\centering
\includegraphics[scale=0.5]{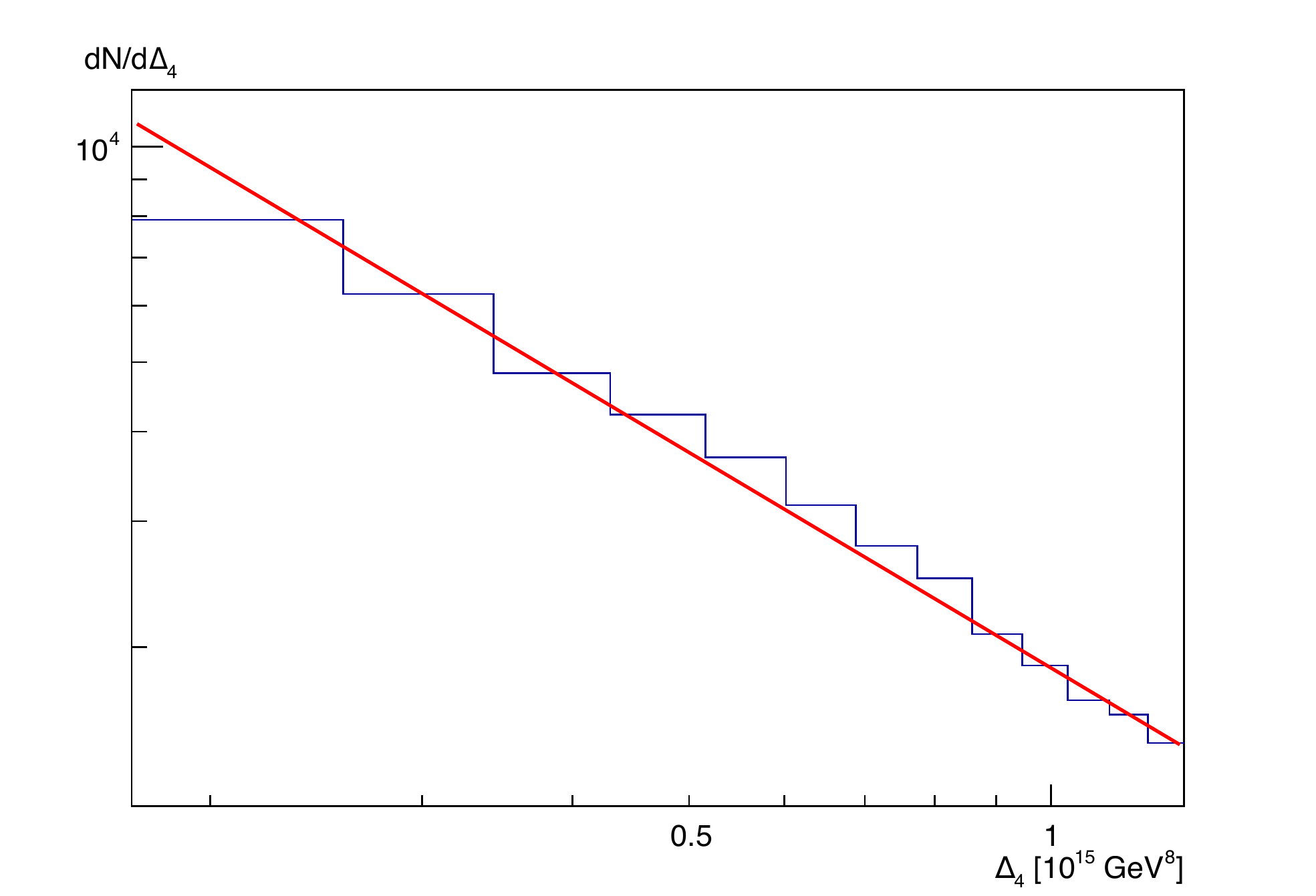}
\caption{\label{5bodyfig}The density of Monte Carlo events arising from a 5-body decay along a direction perpendicular to the boundary of the kinematically accessible region. The red line visually illustrates a scaling law of $\Delta_{4}^{-1}$.}
\end{figure}

In order to isolate the scaling of the volume element near the boundary, an alternative expression can be used~\cite{RevModPhys.36.595}, which locally takes the form:
\beq \label{yb_ps_eqn}
	dPS_n = ({\rm const.}) \times M_X^{-2}\,\left(\prod_{i<j}dm^2_{ij}\right) \Dfour^{-(n-3)/2} \left(\prod_{\alpha\le\beta} \delta(e_{\alpha\beta})\right) \delta\left(\sum_{i<j} m_{ij}^{2} - {\rm const.}\right).
\eeq
Here the $e_{\alpha\beta}$, where $1\le\alpha\le\beta\le n-4$, are a set of $(n-4)(n-3)/2$ constraints that are linear in all $m_{ij}^{2}$ to first order. The form of the $e_{\alpha\beta}$ are complicated, which makes this expression less useful for practical purposes. However since no nontrivial Jacobians arise, the correct scaling $\Dfour^{-(n-3)/2}$ is revealed, which results in a stronger and stronger enhancement near the boundary with increasing $n$. In particular, for $n=5$ the volume element scales as $\Dfour^{-1}$. This can be understood in a similar way to our argument above for $n=4$: as we approach the three dimensional boundary of phase space, a larger number of 4-vectors must become linearly dependent, and the coordinate change from Cartesian coordinates to the $m_{ij}^{2}$ becomes more singular.

We have also verified the scaling for $n=5$ numerically by generating Monte Carlo data for 5-body decays. Specifically, after restricting ourselves to the physical hypersurface specified by the $\Delta_{5}=0$ constraint, we have sampled the density of Monte Carlo events in a narrow tube perpendicular to the boundary near randomly chosen points on the boundary. Our results are shown in figure~\ref{5bodyfig} and they confirm the $\Delta_{4}^{-1}$ scaling, demonstrated visually by the red line.


\section{Mass Determination}
\label{sec:mass}

In this section we will compare the results of mass measurements obtained by using the multi-dimensional nature of the kinematically accessible region in phase space to those obtained from the traditional kinematic edges and endpoints. In order to perform this comparison, we introduce quality-of-fit functions, to be described below, for the two methods, and we search for the spectrum that results in the best fit, using Monte Carlo samples of 100 events each for several decay topologies. By finding the best fit spectrum over many samples, and studying the distribution of the best fit masses, we can evaluate the precision and accuracy of the two techniques. We do this by studying representative benchmark spectra in the decay topologies of interest. Our setup is similar to the analysis performed in ref.~\cite{Agrawal:2013uka}, where final states with three visible particles and one invisible particle were considered. In this paper we extend this to final states with four visible particles and one invisible particle. We use a shorthand notation to classify the topologies of interest by using the multiplicity of final states in each stage of the cascade: for instance ``2+2+3'' denotes a decay topology where the initial state decays through a 2-body decay, the resulting intermediate particle decays through another 2-body decay, and the intermediate particle resulting from the second stage decays through a 3-body decay, where the final state of the last decay stage includes the lightest partner particle.

We do not consider the 2+2+2+2 topology since it has sufficiently many on-shell intermediate particles to be analyzed by polynomial methods. The four on-shell constraints for the intermediate particles together with the 5-body constraint $\Delta_5=0$ are sufficient to restrict the likelihood function to have support on a set of measure zero in the space of mass spectra.  Therefore, the true spectrum can be determined with a finite number of events.
The 2+2+3, 2+3+2 and 3+2+2 topologies are very similar, and therefore we will study the 2+2+3 topology as a representative case. We also study the 3+3 topology which is inequivalent to those. We do not consider decay topologies involving a direct 4-body decay. The endpoint formulas in certain topologies have different analytical forms in different regions of the space of spectra (see appendix~\ref{app:endpoints}). Some forms are functions of mass differences only and cannot contribute to a determination of the overall mass scale, while others do contain some absolute scale dependence. In order to gauge the performance of the endpoint method in a more representative way, we pick two benchmark mass spectra for the 2+2+3 topology. The benchmark mass spectra are listed in table~\ref{spectrum_table}. In particular, benchmark spectrum 2 is expected to be less sensitive to the overall mass scale, as both the $m_{1234}^{2}$ and $m_{234}^{2}$ endpoints depend only on mass differences (see the last lines of equations~\ref{eq:endpoint1234for223} and \ref{eq:endpoint234for223}), while the endpoint formulas for benchmark spectrum 1 do have some dependence on the overall mass scale (line 2 of equations~\ref{eq:endpoint1234for223} and the first line of equation \ref{eq:endpoint234for223}).

\begin{table}[!h]
\centering
\begin{tabular}{| c | c | c | c | c | c |}
\hline
Decay	&	$m_X ~({\rm GeV})$ & $m_Y~({\rm GeV})$ & $m_Z~({\rm GeV})$ & $m_5~({\rm GeV})$ & $m_{1,2,3,4}~({\rm GeV})$ \\ \hline\hline
2+2+3 (1) & 500 & 400 & 150 & 100 & 5\\ \hline
2+2+3 (2) & 400 & 350 & 300 & 100 & 5\\ \hline
3+3 (3) & 500 & 300 & -- & 100 & 5\\ \hline
\end{tabular}
\caption{\label{spectrum_table}Benchmark mass spectra used in our analysis. For the labeling of the masses in the spectra, see figures~\ref{fig:223topology} and \ref{fig:33topology} in appendix~\ref{app:endpoints}.}
\end{table}

The Monte Carlo events are generated using the phase space routines in ROOT~\cite{Brun:1997pa}. We also use the optimization routines in ROOT to find the best fit spectrum. We assume that the underlying decay topology is known; we will comment on the question of determining the decay topology in the next section. We start the optimization procedure within a rectangular box in the space of spectra where each mass is varied by $\pm25\%$ of its correct value (for the multidimensional phase space method) or up to several TeV (for the endpoint method). We perform a random scan inside this box to find the best fit spectrum. We then refine the best fit spectrum using the simulated annealing algorithm.

As mentioned in the introduction, an important caveat in our methods is that all spin correlations are ignored, in other words we use isotropic decays in our Monte Carlo events, and in the quality-of-fit variables described below. Therefore, in the presence of spin correlations, the specific quality-of-fit variable described below for the multidimensional phase space method may develop biases. However, we reiterate that the main purpose of our study in this paper is to provide a proof of principle that multidimensional phase space methods can provide an improvement over kinematic endpoints for mass measurements. Fundamentally, all the information about the spectrum is encoded in the shape of the {\it boundary} of the kinematically accessible region in phase space, not in the distribution of events, which will have additional dependence on the matrix element. The ideal mass measurement analysis would therefore be based on finding the boundary alone, for example by using our methods combined with Voronoi tessellations as has already been done in refs.~\cite{Debnath:2015wra,Debnath:2016mwb,Debnath:2016gwz}. Once the masses have been measured by using the boundary, more sophisticated methods such as matrix element matching can then be utilized for determining the spins of the particles in the decay chain. We proceed with the quality-of-fit variables below mainly to keep the comparison between the two methods as simple as possible for this initial study of the five-body decay chains. We leave a more realistic work incorporating tools such as Voronoi tessellations to future work.

\subsection{Quality-of-fit variable for the kinematic endpoint method}

We define the measured position of a kinematic endpoint as the highest value obtained for the observable in question within the data sample. We construct the quality-of-fit function to quantify the agreement between the measured endpoints and those predicted by the spectrum hypothesis:
\begin{equation}\label{QOFdefn}
	Q =  \Xi \sum_{i=\mathrm{endpoints}}\left( 
		\frac{  \mathcal O_{i}^{\mathrm{predicted}} - \mathcal O_i^{\mathrm{measured}}}{\mathcal O_i^{\mathrm{measured}}}\right)^2 
\end{equation}
where $\Xi = 1$ if all measured endpoints occur at smaller (or equal) values than the predicted ones. If any one of the measured endpoints exceeds the predicted value, the mass hypothesis is rejected ($\Xi$ is taken to be $\infty$). We consider all possible Lorentz-invariant endpoints, with pairs, triplets, etc. of visible final state particles. All endpoints used in our analysis and their predicted values are listed in appendix \ref{app:endpoints}. The best fit mass hypothesis is the one that minimizes $Q$.

\subsection{Quality-of-fit variable for the multidimensional phase space method}

To quantify the quality-of-fit using the multidimensional phase space method, we introduce a likelihood function. In particular, let ${\mathcal L}(\{M_{i}\}|{\rm data})$ denote the likelihood for a hypothesis spectrum $\{M_{i}\}$ given the data. By Bayes' theorem, using a flat prior over spectra, this is proportional to ${\mathcal L}({\rm data}|\{M_{i}\})$, the probability of obtaining the data from the underlying spectrum $\{M_{i}\}$. This probability can be factored over the events in the data sample as
\beq
{\mathcal L}({\rm data}|\{M_{i}\}) = \prod_{\rm events} {\mathcal L}_{\rm event}(\{m_{ij}^{2}\}|\{M_{i}\}),
\eeq
where $\{m_{ij}^{2}\}$ denote all Lorentz-invariant observables in the event. The form of the ${\mathcal L}_{\rm event}$ factors and the details of their calculation are described in appendix~\ref{app:boundarylikelihood}. Ultimately, we bring the likelihood functions for each topology into a standard form
\beq
\mathcal{L}_{\rm event} = \Theta[\mathcal{D}_1] \cdots \Theta[\mathcal{D}_m] \times \mathcal{N} \times \mathcal{F}
\eeq
where for any given decay topology, the $\Theta[\mathcal{D}]$ factors encode the kinematically accessible region in phase space, $\mathcal{F}$ contains all dependence on the hypothesis spectrum $\{M_{i}\}$, and $\mathcal{N}$ includes all remaining dependence on the observables in the events. Note that as in the setup for the kinematic endpoint method, spectra for which there exist events that fall outside the (hypothetical) kinematically accessible region are considered excluded. Since the phase space density becomes large near the boundary of the kinematically accessible region, the likelihood function favors spectra where as many events as possible lie near the boundary, with no events lying outside the boundary.  The best fit mass hypothesis is the one that maximizes ${\mathcal L}$ (to be more precise, we use the logarithm of ${\mathcal L}$).

\subsection{Analysis and Results}
As mentioned above, kinematic endpoint methods are generically much more sensitive to mass differences in the spectrum than to the overall mass scale, parameterized e.g. by the mass of the lightest partner. Therefore, when the statistical distribution of best fit values for the mass of any particle in the spectrum is considered, the spread in the distribution is dominated by the uncertainty in the overall scale. In order to better compare the performance of the two methods to the overall mass scale and to the mass gaps in the spectrum separately, it is preferable to find an alternative parametrization for spectra rather than using the masses of the individual particles. In particular, we parameterize the spectrum in terms of one parameter that sets the overall mass scale, and three other parameters that only depend on mass gaps.

\begin{figure}
\centering
\includegraphics[scale=0.46]{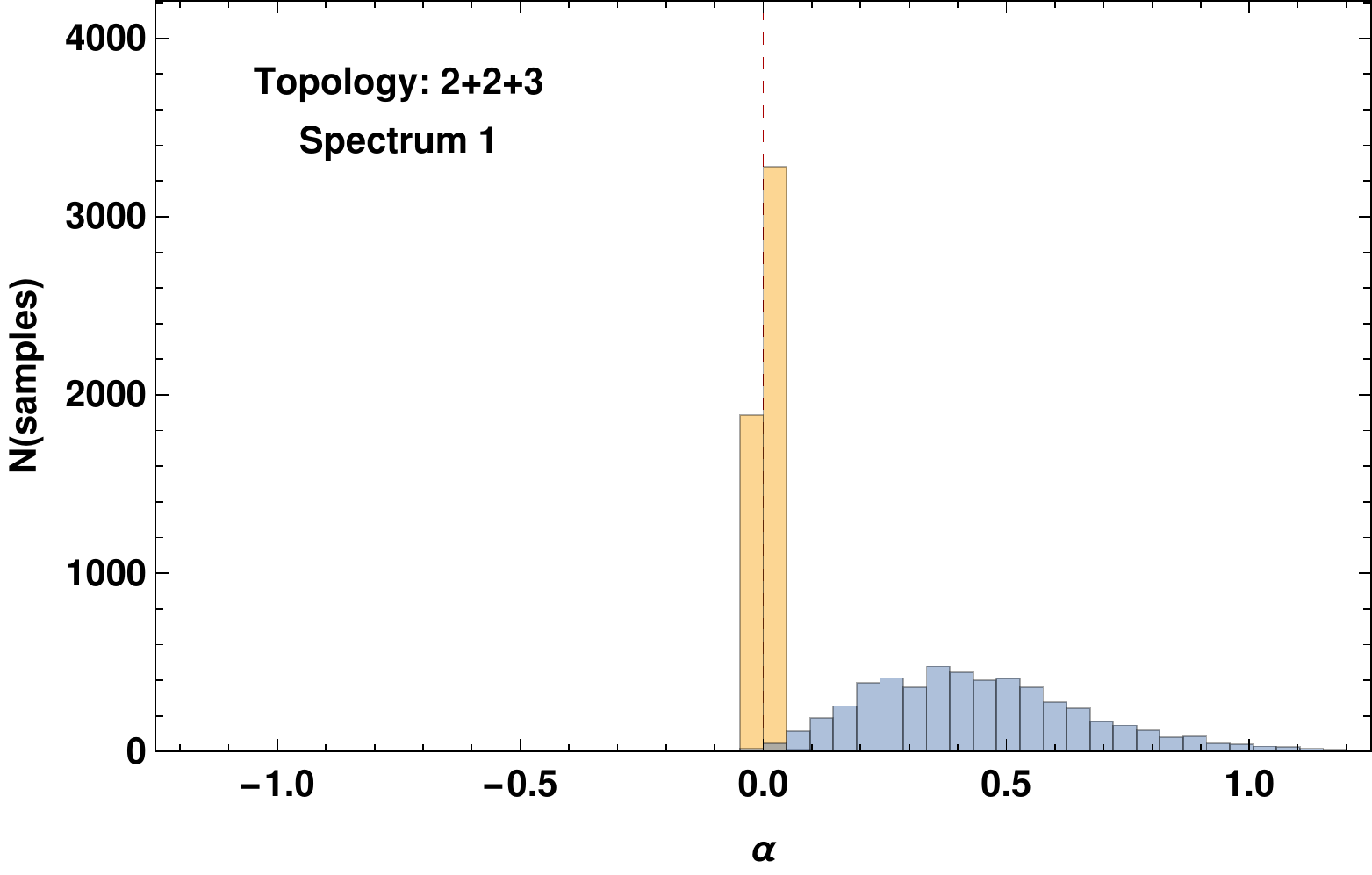}
\includegraphics[scale=0.46]{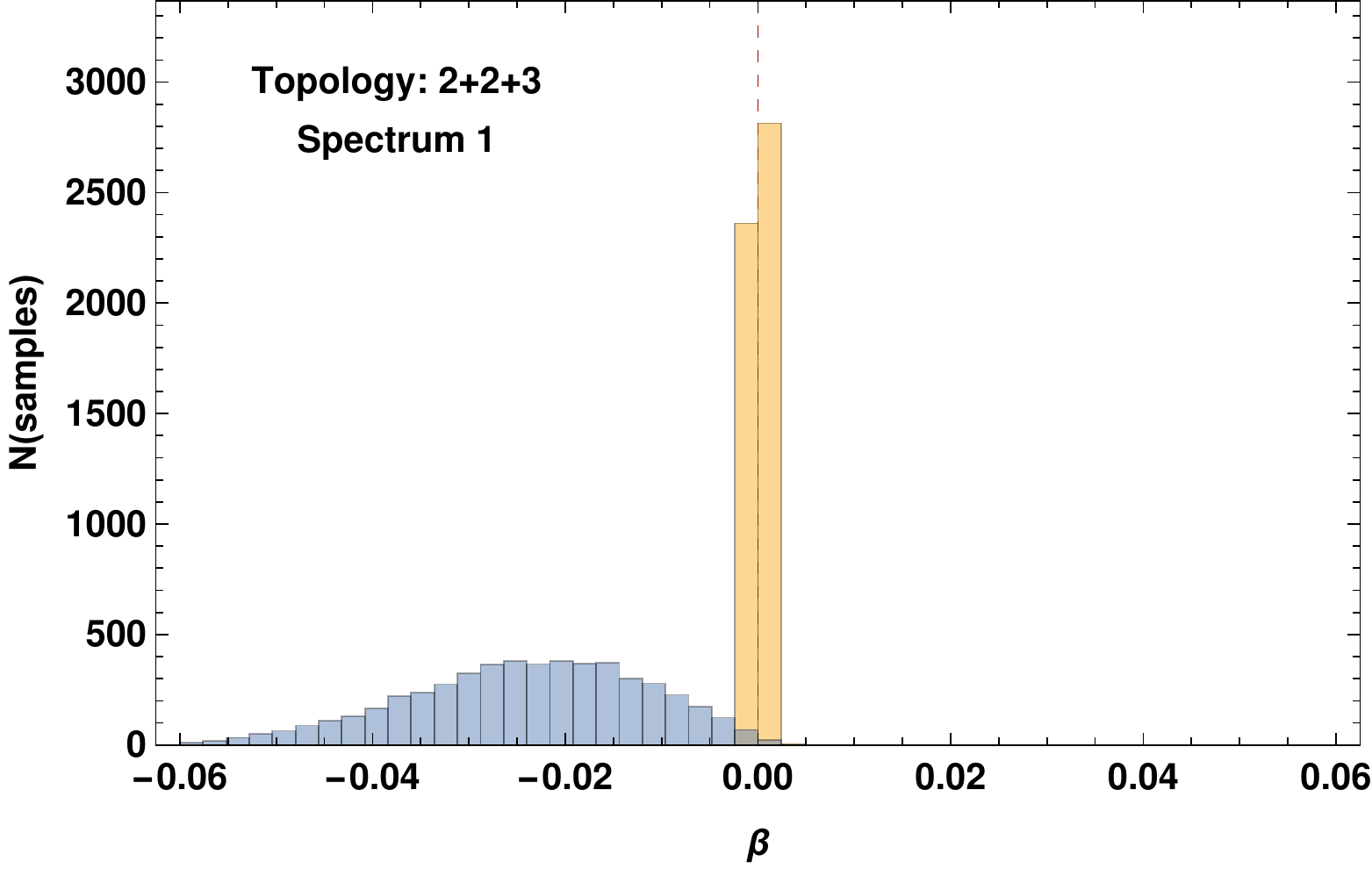} \\
\includegraphics[scale=0.46]{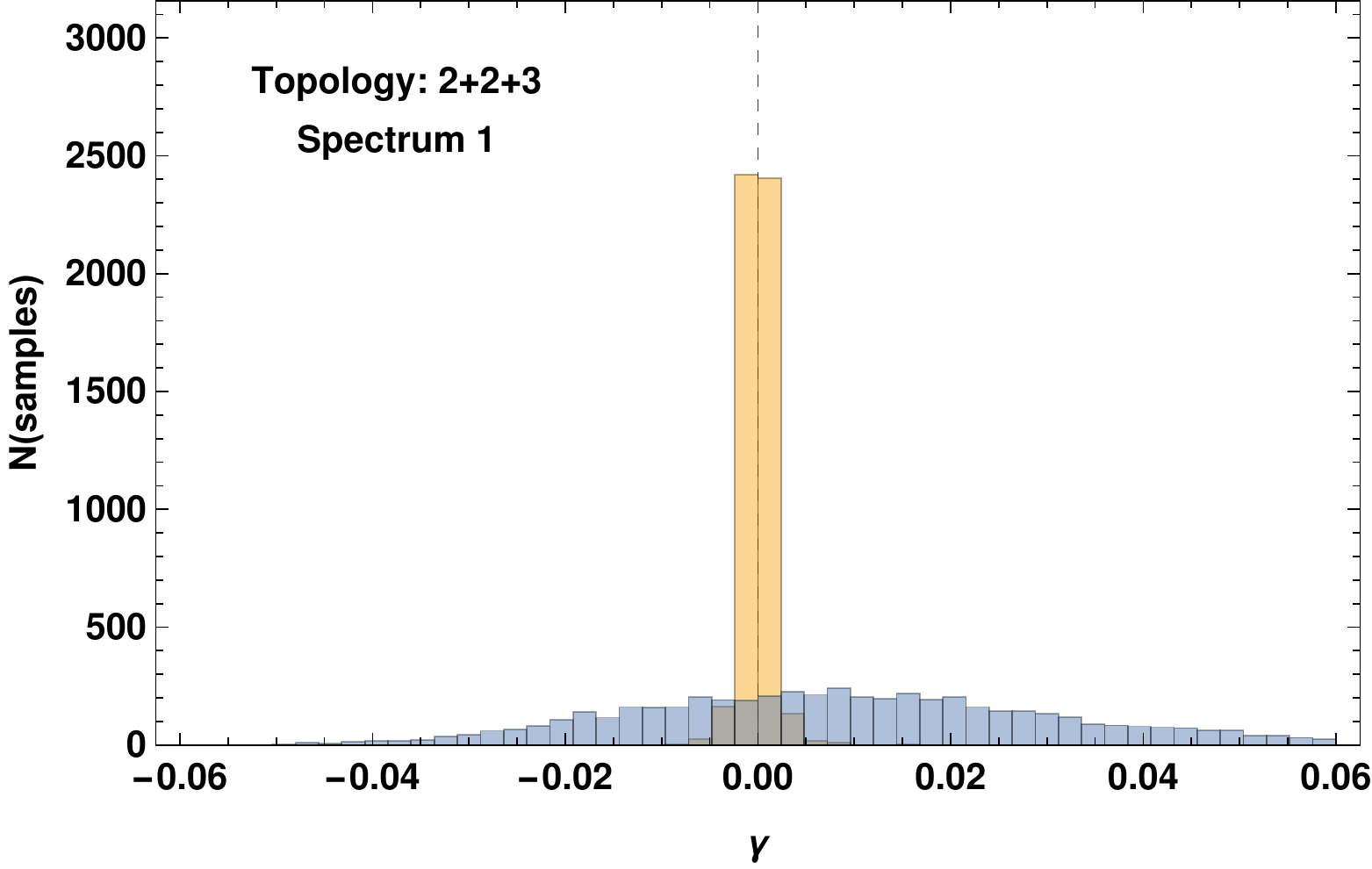}
\includegraphics[scale=0.46]{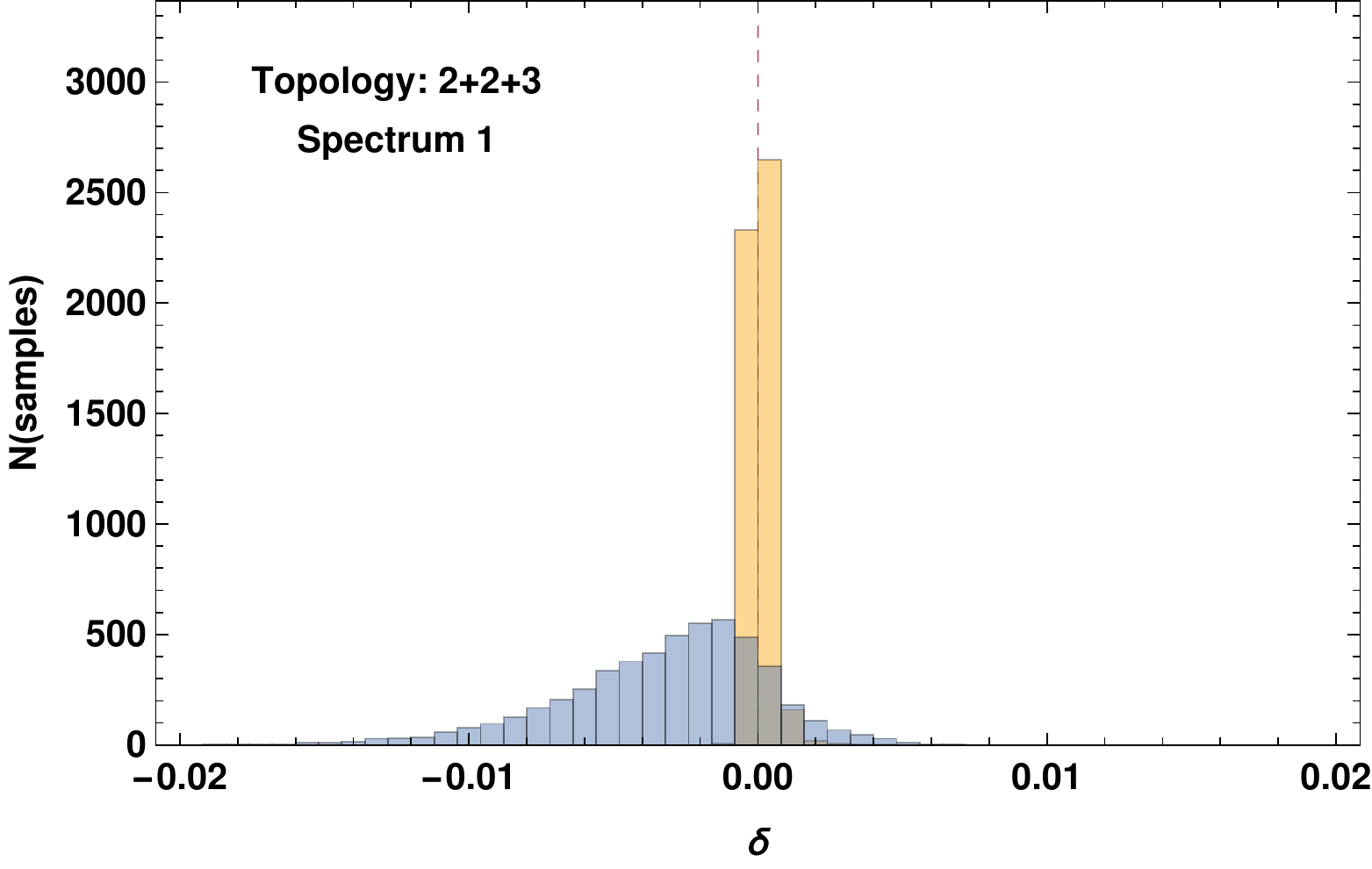}
\caption{\label{223_mass_fig_spectrum_1}Distribution of the best fit values of $\alpha$, $\beta$, $\gamma$ and $\delta$ (defined in equation~\ref{eq:alpha223}) for the kinematic endpoint method (blue) and the multidimensional phase space method (yellow), using the first benchmark spectrum for the 2+2+3 topology and data samples of 100 events. The true masses correspond to $\alpha$, $\beta$, $\gamma$ and $\delta$ all being zero.}
\end{figure}

For the 2+2+3 topology, we define the dimensionless parameters $\{\alpha, \beta, \gamma,\delta\}$ as  
\begin{equation}\label{eq:alpha223}
	M_{i} =M_{i}^{\rm true}+ \left(\alpha\, V_\alpha  + \beta\, V_\beta  + \gamma\, V_\gamma  + \delta\, V_\delta \right)_{i}\times(100~{\rm GeV}),
\end{equation}
where $M_{1}$ parameterizes the (hypothetical) mass of the initial decaying particle, $M_{4}$ parameterizes the (hypothetical) mass of the lightest partner, $M_{i}^{\rm true}$ denote the benchmark mass values that were used to generate the Monte Carlo events, and the vectors $V$ are defined as
\begin{eqnarray}
V_\alpha &= \{1,1,1,1\}, \\
V_\beta &= \{1,-1,0,0\},\\
V_\gamma &= \{1,1,-1,-1 \}, \\
V_\delta &= \{0,0,1,-1 \}.
\end{eqnarray}
Thus the coordinate $\alpha$ parametrizes the overall mass scale. The allowed range of $\alpha$, $\beta$, $\gamma$ and $\delta$ are chosen such that the hierarchy of masses is preserved, and all masses remain positive.

Similarly, for the 3+3 topology we define $\{\alpha, \beta, \gamma\}$ as  
\begin{equation}\label{eq:alpha33}
	M_{i} = M_{i}^{\rm true} + \left(\alpha\, V_\alpha  + \beta\, V_\beta  + \gamma\, V_\gamma  \right)_{i}\times(100~{\rm GeV}),
\end{equation}
where
\begin{eqnarray}
V_\alpha &= \{1,1,1\} \\
V_\beta &= \{0,1,-1\}\\
V_\gamma &= \{2,-1,-1 \}.
\end{eqnarray}
Again, $\alpha$ parameterizes the overall mass scale, and similar consideration as above apply in choosing the allowed range for these parameters.

\begin{figure}
\centering
\includegraphics[scale=0.46]{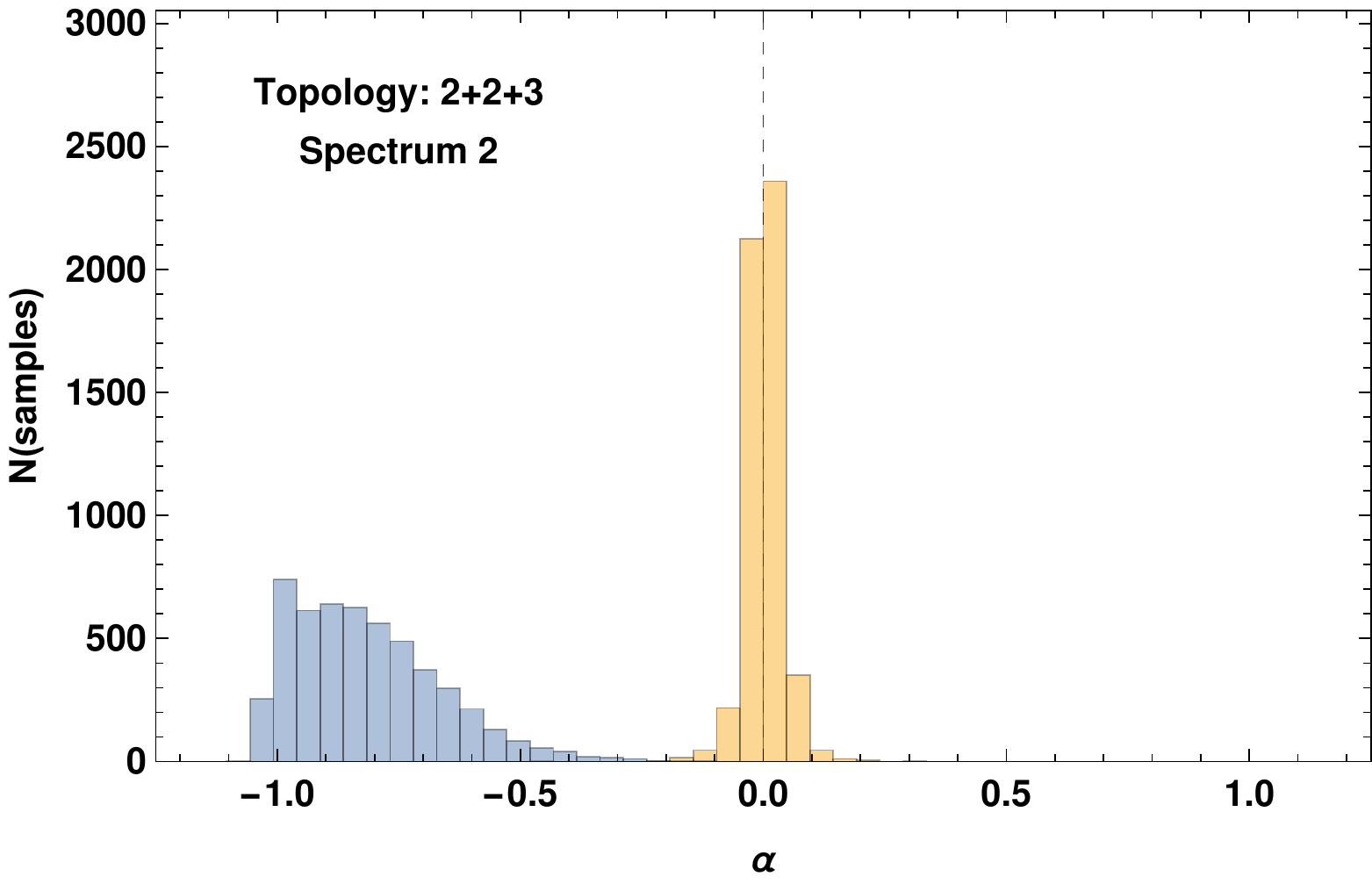}
\includegraphics[scale=0.46]{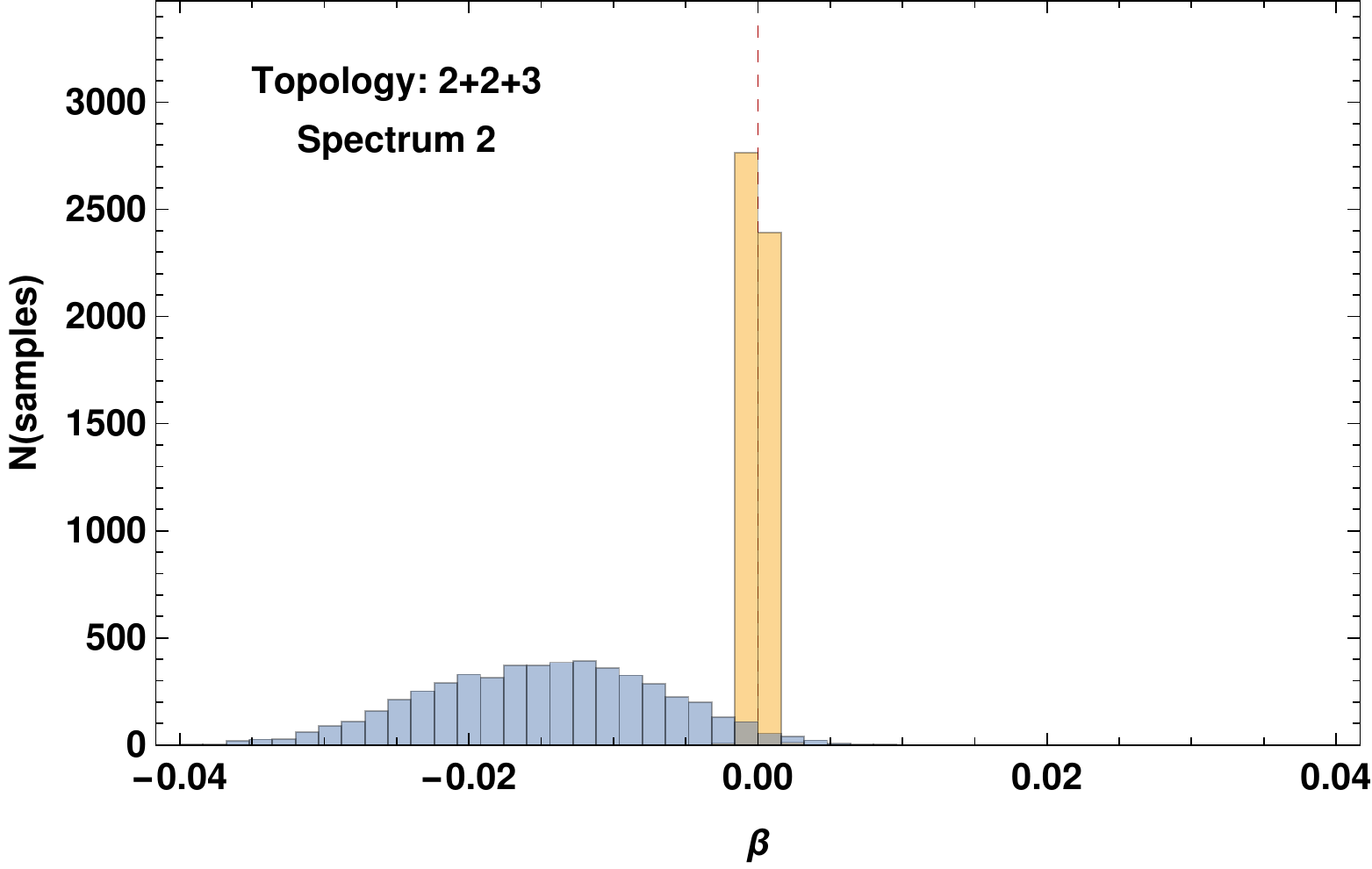} \\
\includegraphics[scale=0.46]{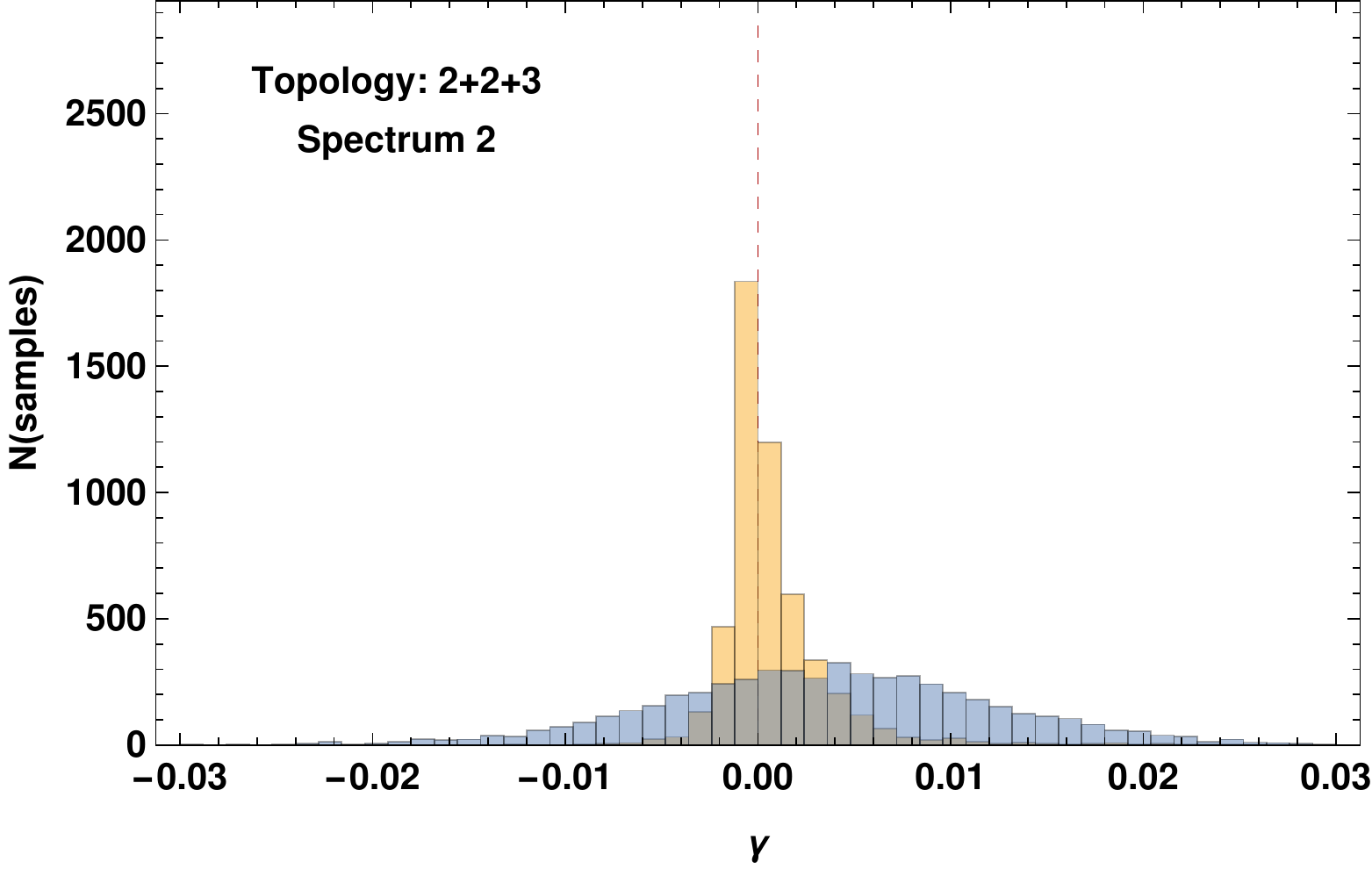}
\includegraphics[scale=0.46]{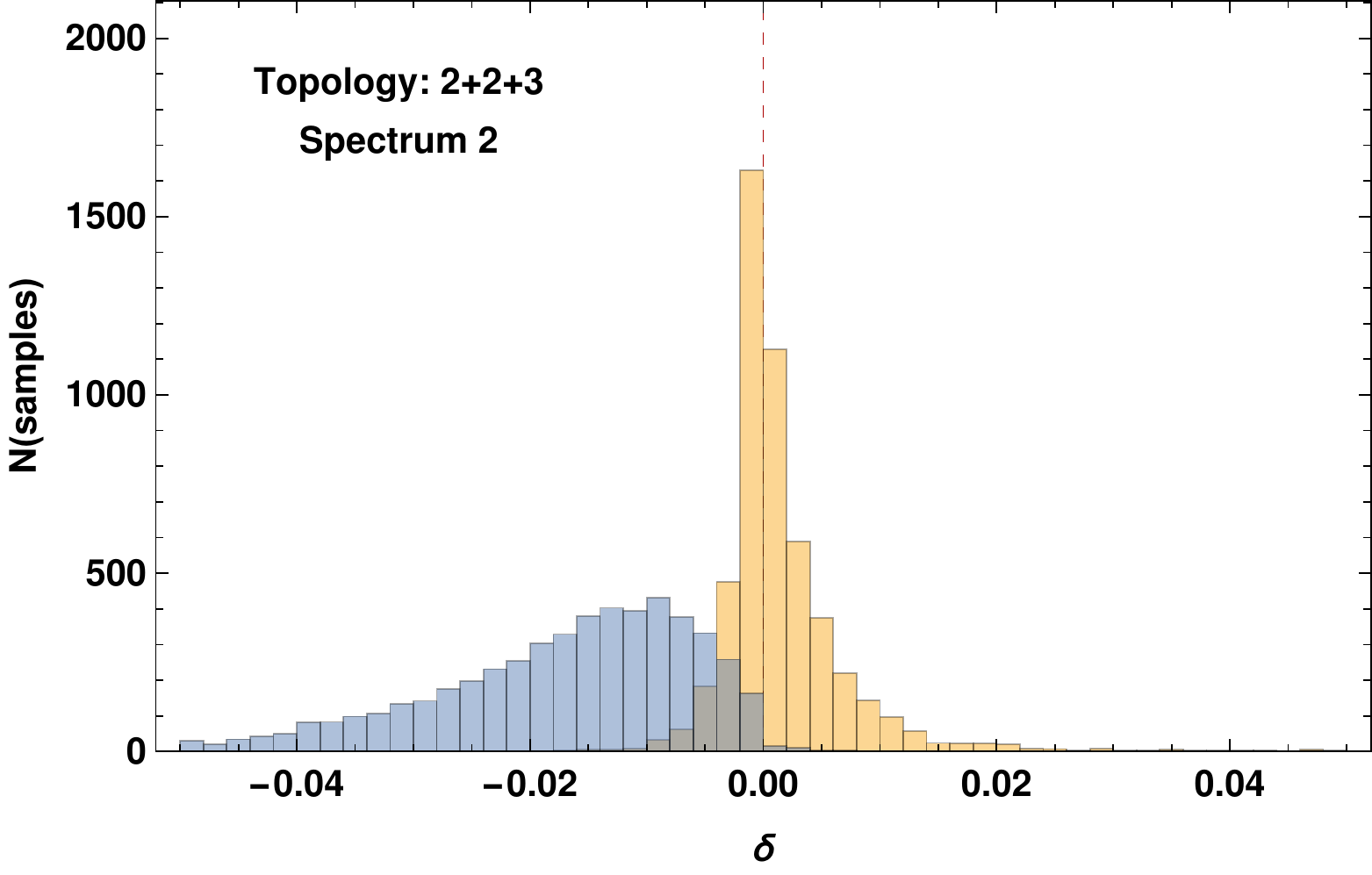}
\caption{\label{223_mass_fig_spectrum_2}Distribution of the best fit values of $\alpha$, $\beta$, $\gamma$ and $\delta$ (defined in equation~\ref{eq:alpha223}) for the kinematic endpoint method (blue) and the multidimensional phase space method (yellow), using the second benchmark spectrum for the 2+2+3 topology and data samples of 100 events. The true masses correspond to $\alpha$, $\beta$, $\gamma$ and $\delta$ all being zero.}
\end{figure}

\begin{figure}
\centering
\includegraphics[scale=0.46]{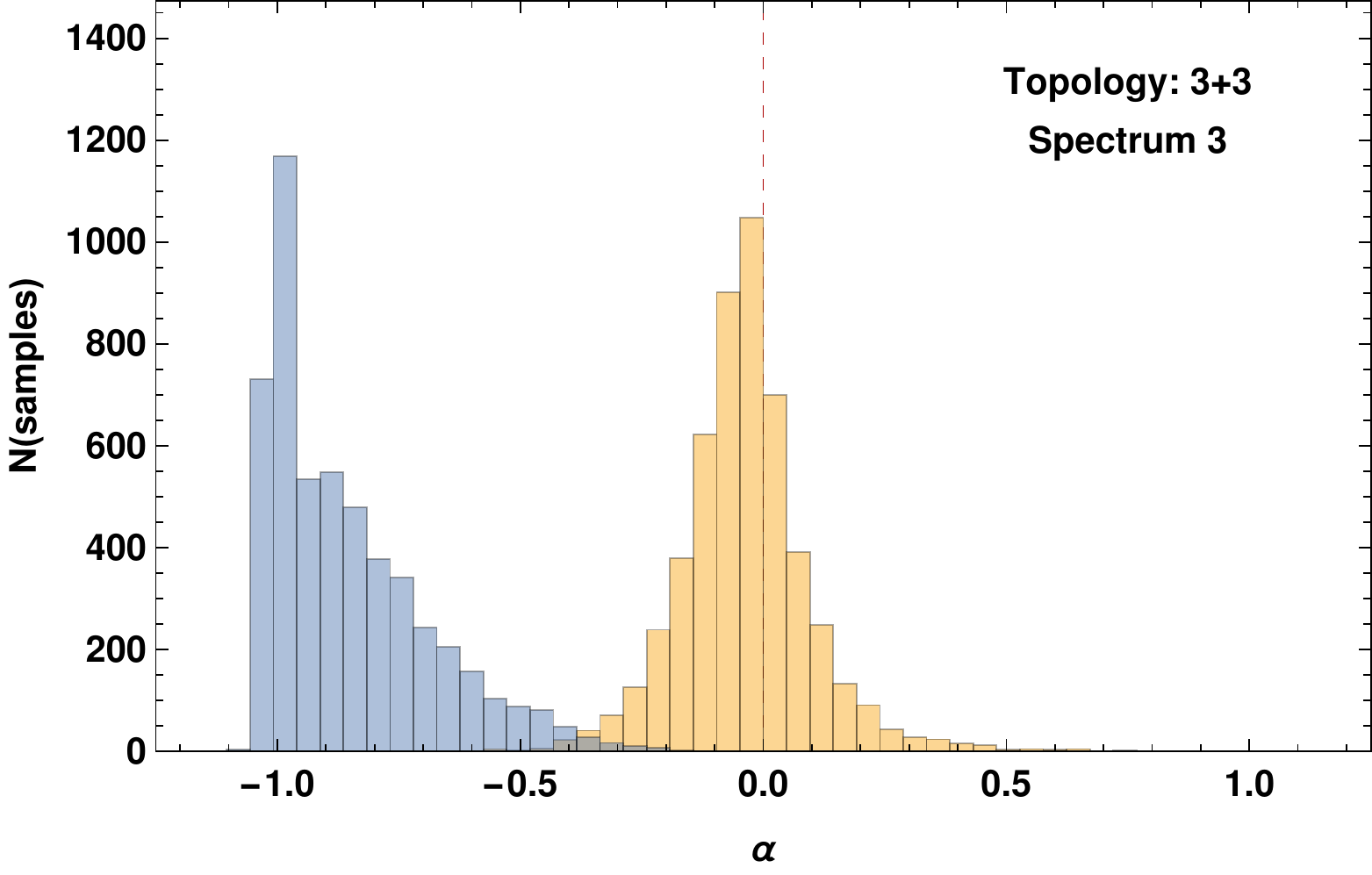}
\includegraphics[scale=0.46]{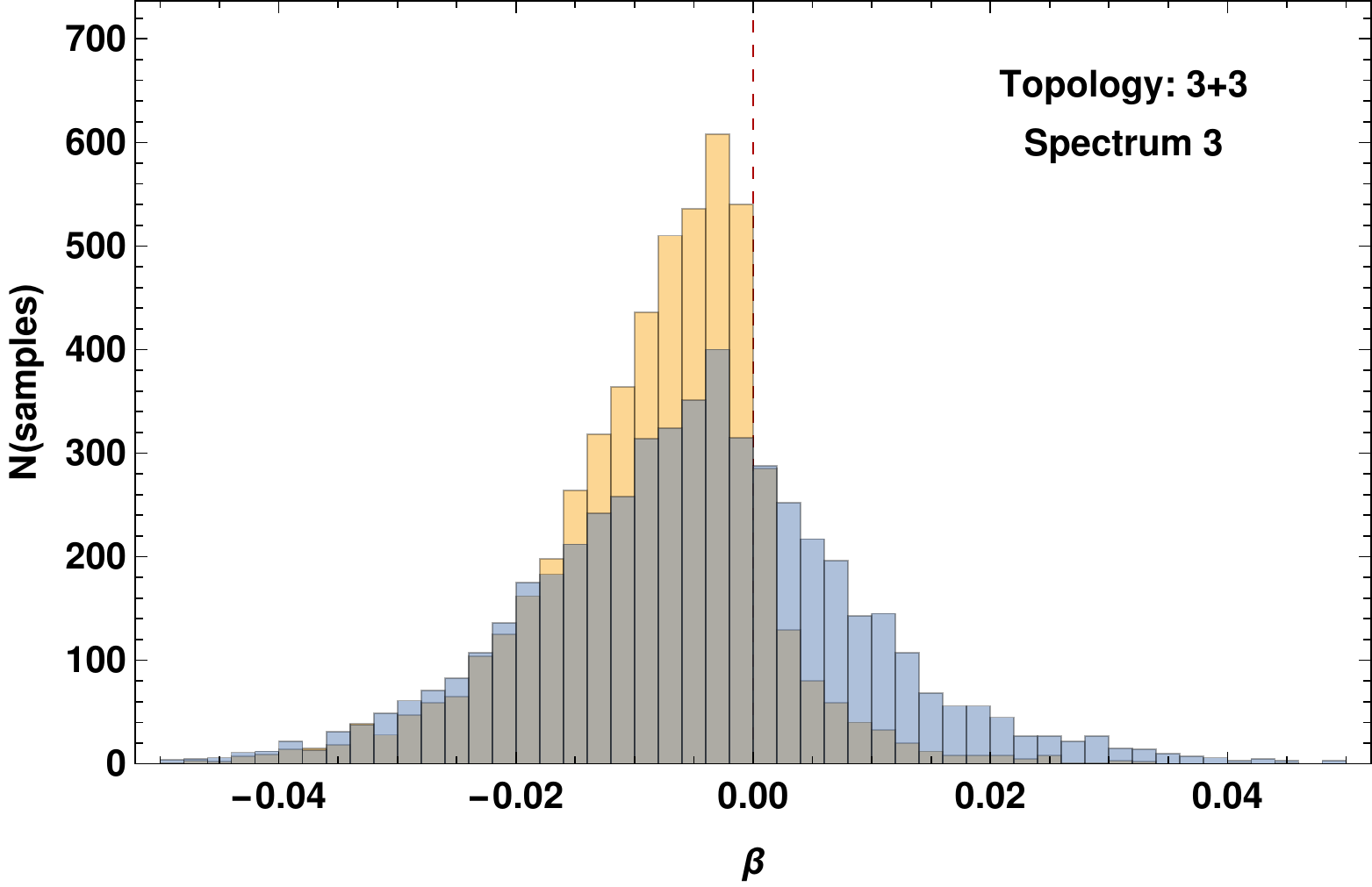}
\includegraphics[scale=0.46]{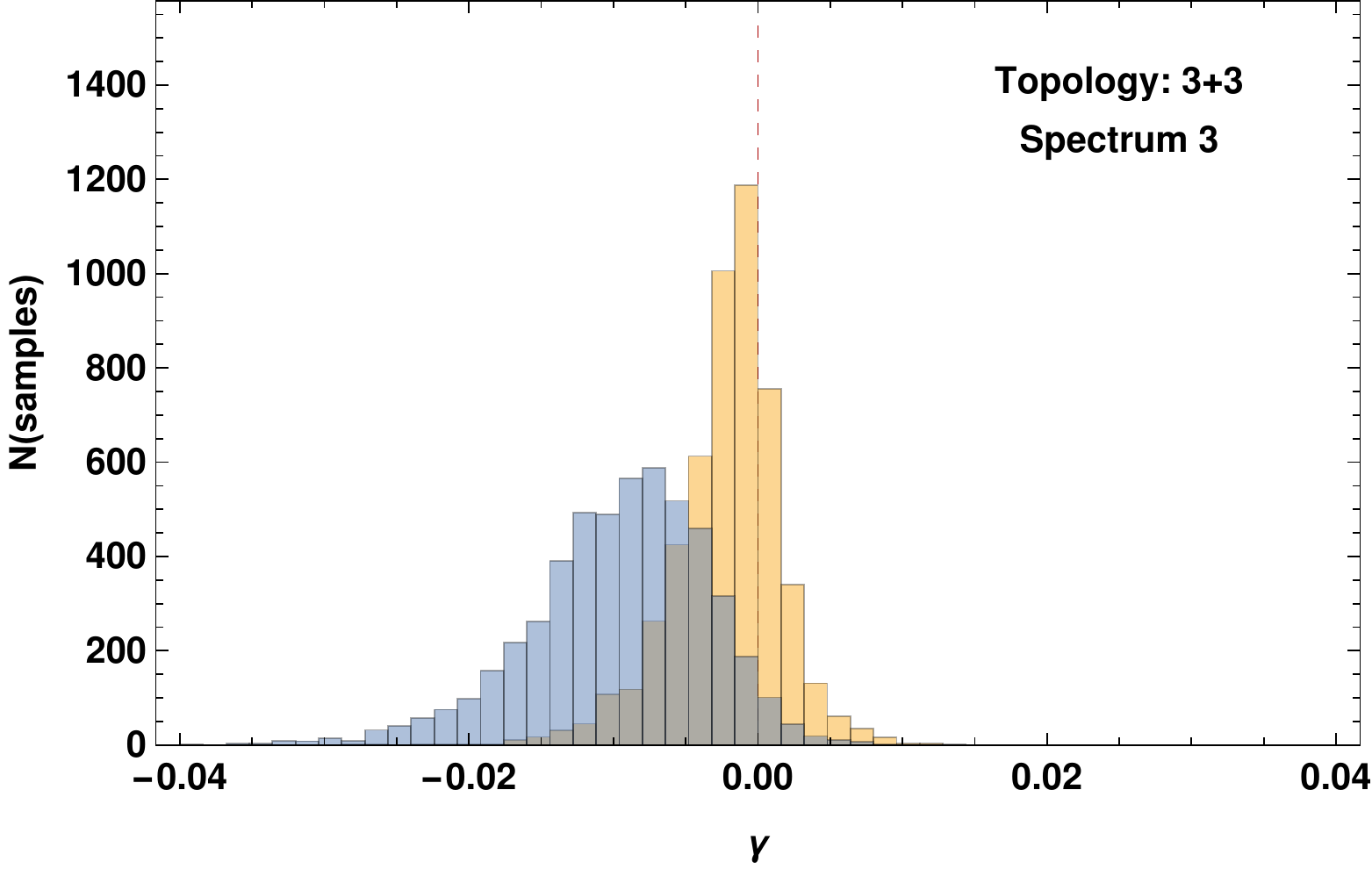}
\caption{\label{33_mass_fig}Distribution of the best fit values of $\alpha$, $\beta$ and $\gamma$ (defined in equation~\ref{eq:alpha33}) for the kinematic endpoint method (blue) and the multidimensional phase space method (yellow), using the benchmark spectrum for the 3+3 topology and data samples of 100 events. The true masses correspond to $\alpha$, $\beta$ and $\gamma$ all being zero.}
\end{figure}

Our results for the 2+2+3 topology are shown in figure~\ref{223_mass_fig_spectrum_1} for benchmark spectrum 1, in figure~\ref{223_mass_fig_spectrum_2} for benchmark spectrum 2. The results for the 3+3 topology is shown in figure~\ref{33_mass_fig}. The mean value and standard deviation of the distributions for $\alpha$, $\beta$, $\gamma$ and $\delta$ are listed in table~\ref{table:mass_results}. It is easy to see that the conclusions obtained for the four-body decay topologies~\cite{Agrawal:2013uka} continue to hold, namely that the multidimensional phase space method yields both more precise and more accurate results for the overall mass scale as well as for the mass gaps. The reasons for the mean values of the distributions obtained by the kinematic endpoint method to be biased away from the correct masses is similar to those discussed in appendix C of ref.~\cite{Agrawal:2013uka} for the four-body final states. 
Note also that for the 2+2+3 topology, although the $\alpha$ distribution for the kinematic endpoint method of benchmark spectrum 1, which was chosen to have {\it lesser} sensitivity on the overall scale, seems to be broader compared to benchmark spectrum 2, this is somewhat misleading. The lower end of the $\alpha$ distribution for benchmark spectrum 2 is cut off by the constraint that all masses in the spectrum be positive numbers, which obscures the true spread in the distribution.

\begin{table}
\centering
	\begin{tabular}{| >{$}c<{$} | >{$}c<{$} | >{$}c<{$} |}
	\hline
	\mathrm{} & \mathrm{Multidimensional~Phase~Space} & \mathrm{Kinematic~Endpoints} \\ \hline
	\multicolumn{3}{|c|}{2+2+3 Topology Benchmark Spectrum 1}\\ \hline
		m_X~({\rm GeV}) & 500 \pm 1 & 543\pm 24\\
		m_Y~({\rm GeV}) & 400 \pm 1 & 447\pm 26\\
		m_Z~({\rm GeV}) & 150 \pm 1 & 193\pm 22\\
		m_5~({\rm GeV}) & 100 \pm 1 & 143\pm 22\\ \hline
	\alpha & (0.2\pm 0.8)\times 10^{-2} & 0.4\pm 0.2 \\ 
	\beta & (0.04\pm 0.3)\times 10^{-3} & (-2\pm 1)\times 10^{-2}\\
	\gamma & (0.007 \pm 1.7)\times 10^{-3} & (1 \pm 2)\times 10^{-2}\\ 
	\delta & (0.1\pm 0.8)\times 10^{-3} & (-0.3 \pm 0.4)\times 10^{-2}\\ \hline
	\multicolumn{3}{|c|}{2+2+3 Topology Benchmark Spectrum 2}\\ \hline
		m_X~({\rm GeV}) & 400 \pm 4 & 317 \pm 15\\
		m_Y~({\rm GeV}) & 350 \pm 4 & 270 \pm 14\\
		m_Z~({\rm GeV})  & 300 \pm 4 & 216 \pm 15\\
		m_5~({\rm GeV}) & 100 \pm 5 & 20 \pm 16\\ \hline
	\alpha & (0.3 \pm 4)\times 10^{-2} & -0.8 \pm 0.2 \\ 
	\beta & (-0.004\pm 0.5)\times 10^{-3} & (-1\pm 0.8)\times 10^{-2} \\
	\gamma & (0.8 \pm 4)\times 10^{-3} & (0.4\pm 0.9)\times 10^{-2} \\ 
	\delta & (2 \pm 7) \times 10^{-3} & (-2 \pm 1)\times 10^{-2} \\ \hline
	\multicolumn{3}{|c|}{3+3 Topology}\\ \hline
		m_X~({\rm GeV}) & 496 \pm 13 & 413 \pm 16 \\
		m_Y~({\rm GeV}) &  296 \pm 14 & 215 \pm 17 \\
		m_5~({\rm GeV}) & 98 \pm 15 & 16 \pm 17 \\ \hline
	\alpha & -0.04 \pm 0.1 & -0.9\pm 0.2\\ 
	\beta & (-0.8 \pm 1) \times 10^{-2} & (-0.5\pm 1.4)\times 10^{-2}\\
	\gamma & (-2 \pm 4) \times 10^{-3} & (-1 \pm 0.6)\times 10^{-2}\\ 
	\hline	
	\end{tabular}	
\caption{\label{table:mass_results}The mean value and standard deviation of the distributions masses in the spectrum as well as of the parameters $\alpha$, $\beta$ etc. for the two benchmark spectra of the 2+2+3 topology, and for the benchmark spectrum of the 3+3 topology, for data samples of 100 events.}
\end{table}


\section{Topology Determination}
\label{sec:topology}

In this section we will consider how different event topologies may be distinguished from one another by using the distribution of events in phase space. We will consider both 4-body and 5-body final states, since ref.~\cite{Agrawal:2013uka} did not consider the question of topology determination. In particular, let $\{T_{i}\}$ be the set of event topologies that are compatible with the number of observed particles, with one invisible particle assumed to be produced in the last stage of the cascade. We will now write the likelihood function as ${\mathcal L}(T_{i},\{M_{i}\}|{\rm data})$, making the dependence on the topology explicit. As before, with a flat prior over topologies and spectra, the likelihood can be related to the probability of obtaining a given distribution of events from an underlying topology
\begin{equation}
{\mathcal L}(T_{i},\{M_{i}\}|{\rm data})\propto {\mathcal L}({\rm data}|T_{i},\{M_{i}\}).
\end{equation}

We can now use the likelihood functions listed in appendix~\ref{app:boundarylikelihood}, in the standard form
\begin{equation}
{\mathcal L}({\rm data}|T_{i},\{M_{i}\})=\prod_{\rm events} \Theta[\mathcal{D}_1] \cdots \Theta[\mathcal{D}_m] \times \mathcal{N}(\{m_{ij}^{2}\}) \times \mathcal{F}(\{m_{ij}^{2}\},\{M_{i}\}).
\end{equation}
As for the analysis for mass measurements, we adopt $\log{\mathcal L}$ as the quality of fit variable. Maximizing over spectra as before, statistical statements (such as exclusion with a given confidence level) can then be made about a number of possible topology hypotheses based on the data. We will not attempt to perform a detailed analysis of this type in this work, since the idealizations we work with, such as perfect energy resolution and the absence of backgrounds and combinatoric effects, would render the conclusions unreliable.

\begin{figure}
\centering
\includegraphics[scale=0.5]{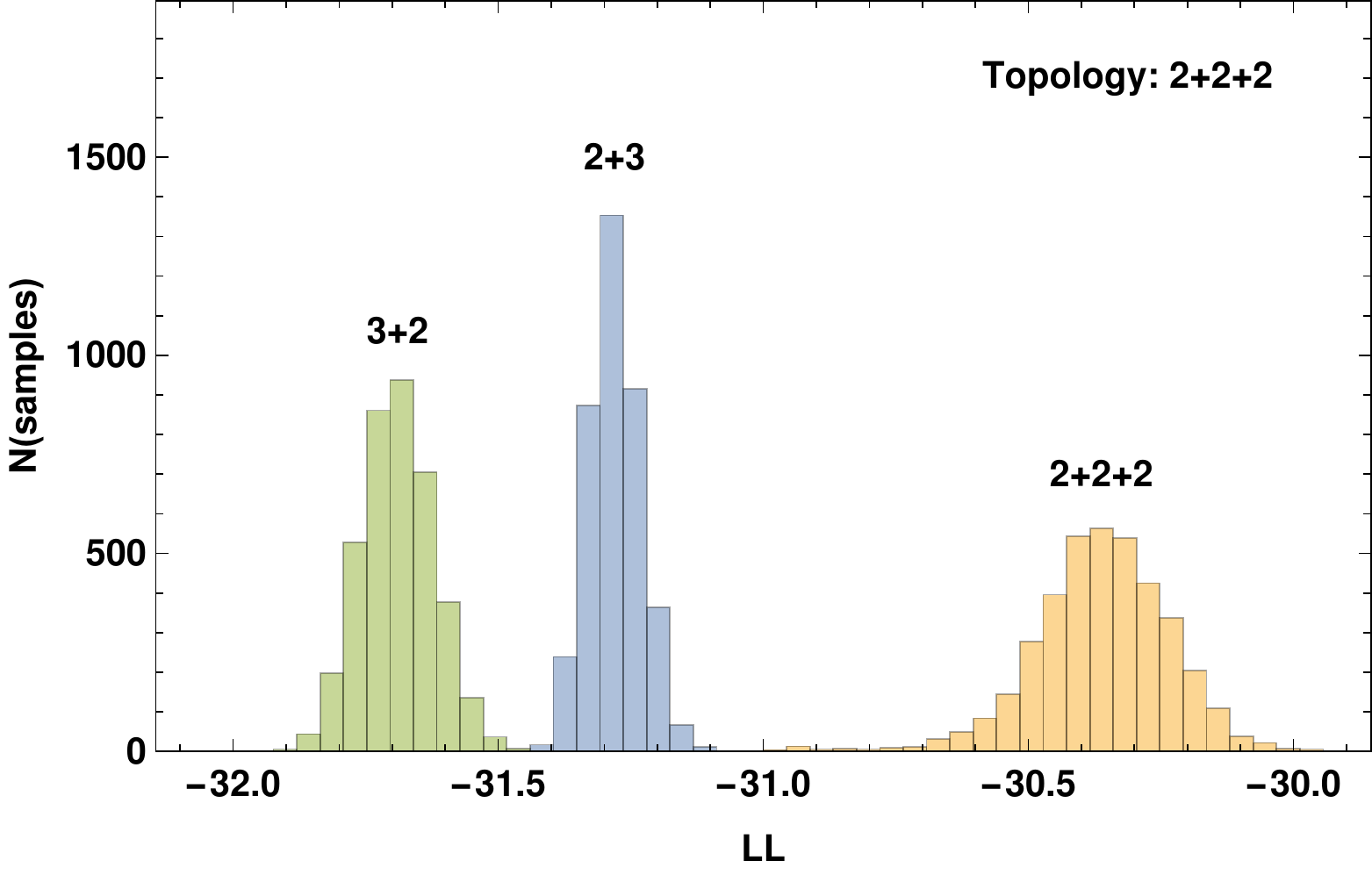}
\caption{\label{fig:topo222}For data samples of 100 events each generated with the spectrum $(500,350,200,100)~$GeV, the distribution of log-likelihood values for the $2+2+2$, $2+3$ and $3+2$ topology hypotheses where the likelihood is maximized over spectra for each sample and each topology hypothesis.}
\end{figure}

Nevertheless, one general conclusion can be drawn immediately: When a topology hypothesis $\tilde{T}$ contains more on-shell particles than the ``true'' topology $T$, it can be ruled out (for any spectrum) with a very small number of events. 
Indeed, for the hypothesis $\tilde T$, the optimization over mass spectra will be trying to enforce an on-shell constraint among the visible particles where no such constraint is actually obeyed by the data. In general, there is no reason for a constraint that appears to be satisfied by one event to also be satisfied by any other. 
Conversely, a choice for $\tilde{T}$ that contains fewer on-shell particles than $T$, while it cannot be ruled out completely, will generally result in a significantly lower likelihood than when the correct topology hypothesis is used, since $\tilde{T}$ will not provide a very good fit to the distribution of events in the data.

Let us demonstrate this on a specific example. The $2+2+2$ topology with the spectrum $(500,350,200,100)~$GeV was used to generate Monte Carlo samples of 100 events each, and for the topology hypotheses $2+2+2$, $2+3$, $3+2$ and $4$, all possible spectra were scanned until the spectrum with the highest likelihood was found for each sample. Note that unlike in the analysis in section~\ref{sec:mass}, for an incorrect topology hypothesis there is no ``correct'' mass point to center the scan region on, therefore we scan the spectra over a larger region where each mass is varied between zero and several TeV. The distribution of the best-fit log-likelihood over samples for each topology hypothesis is shown in figure~\ref{fig:topo222}. In accordance with our expectations, the $2+3$ and $3+2$ topologies with fewer on-shell particles result in a poor fit, and the correct topology results in the highest likelihoods.

It should be noted that for certain incorrect hypotheses, there exists a runaway direction in the space of spectra $\{M_{i}\}$, namely the likelihood increases as all masses go to infinity with fixed mass gaps. This happens for instance when a direct 4-body decay topology hypothesis is used in the example above. In addition to being completely unphysical (which is why they are not plotted in figure~\ref{fig:topo222}), the likelihood values for this topology hypothesis in any case turn out to be smaller than for the other topology hypotheses. Runaway directions do not exist for the correct topology hypothesis, and therefore the presence of a runaway direction can be used to rule out a topology hypothesis.

Based on the above considerations, a rather general conclusion can be reached that when analyzing a given data sample, the correct topology is among those hypotheses that have the highest number of on-shell particles and that are not immediately ruled out. If there is only one such hypothesis ($2+2+2$ in the above example), then it must be the correct one. Things are more interesting when there are competing hypotheses with the same number of on-shell particles.

\begin{figure}
\centering
\includegraphics[scale=0.5]{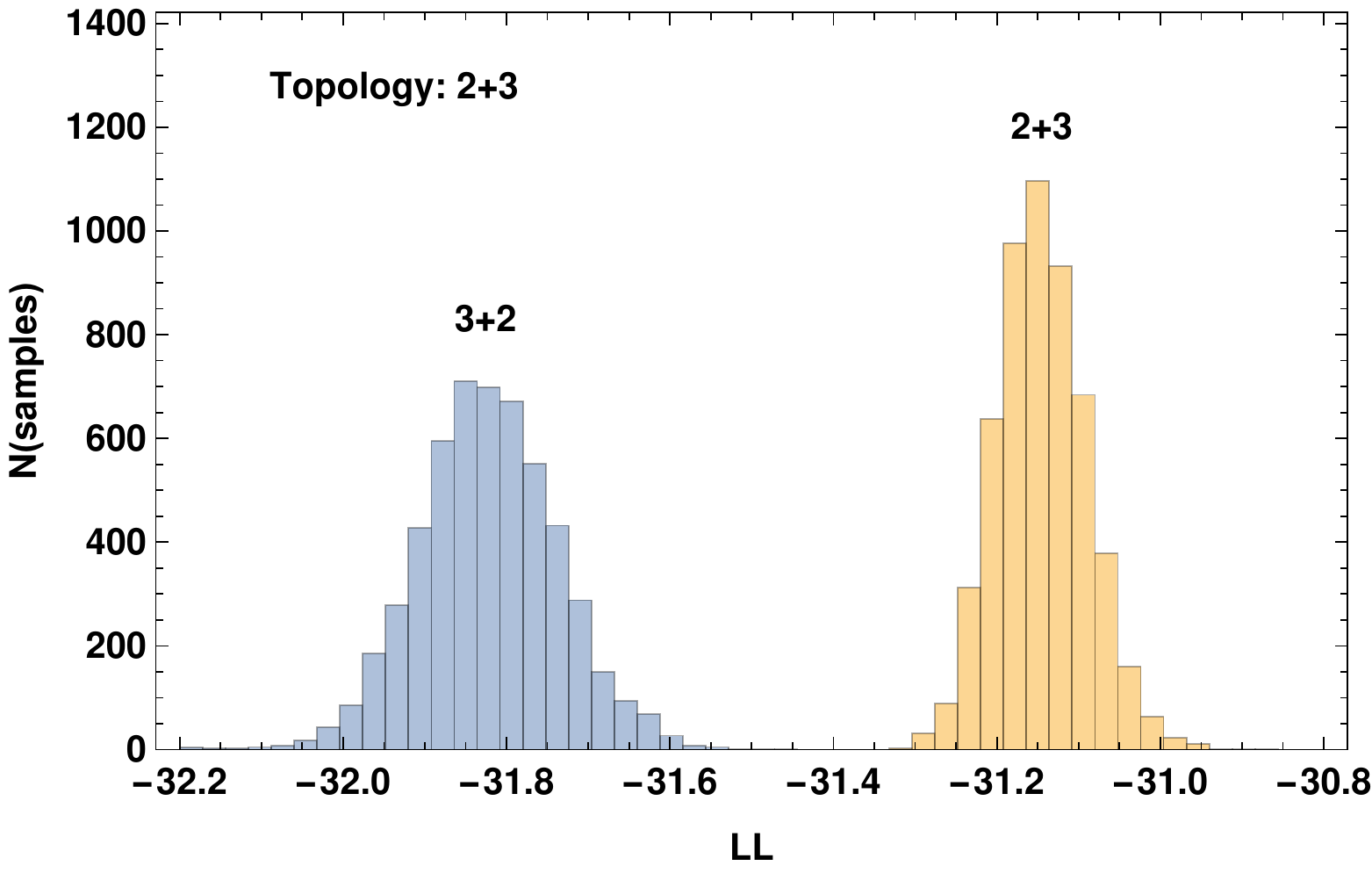}
\caption{\label{fig:topo23}For data samples of 100 events each generated with the spectrum $(500,350,100)~$GeV and the $2+3$ hypothesis, the distribution of log-likelihood values for the $2+3$ and $3+2$ topology hypotheses where the likelihood is maximized over spectra for each sample and each topology hypothesis.}
\end{figure}

For the final states with three visible particles and one invisible particle, the following outcomes are therefore possible:
\begin{itemize}
{\item The data does not rule out the $2+2+2$ topology hypothesis, which is then established as the correct one.}
{\item The data is not compatible with the $2+2+2$ topology hypothesis but it is compatible with the $2+3$ and $3+2$ topology hypotheses. This is the only nontrivial case that can arise with this number of final state particles, and a statistical analysis would be needed to find the best fit topology hypothesis. An example of this is demonstrated in figure~\ref{fig:topo23}, where the log-likelihood distributions are plotted for the two competing hypotheses for data samples of 100 events each, generated with the $2+3$ topology and the spectrum $(500,350,100)~$GeV. The log-likelihood distribution clearly favors the correct topology.}
{\item The data is only compatible with a direct 4-body decay hypothesis, which is then established as the correct topology.}
\end{itemize}

Similarly, for the final states with four visible particles and one invisible particle, the possible outcomes are:
\begin{itemize}
{\item The data is compatible with the $2+2+2+2$ topology hypothesis, which consequently must be the correct one.}
{\item The data rules out the $2+2+2+2$ topology hypothesis but it is compatible with the $2+2+3$, $2+3+2$ and $3+2+2$ topology hypotheses. Since these have the same number of on-shell particles, a statistical analysis would need to be performed to determine the correct topology hypothesis. We have performed a numerical study of this scenario with samples of 100 events each, generated with the $2+2+3$ topology and the spectrum $(500,350,200,100)~$GeV. The log-likelihood distribution not only favors the correct topology but in fact the incorrect topology hypotheses are ruled out completely since no spectrum can be found that is consistent with the data.}
{\item If the data is not compatible with any of the above, then the $3+3$ topology hypothesis is the most likely fit, though technically $4+2$ or $2+4$ are also potential topology hypotheses since they have the same number of on-shell intermediate particles. It is rare for particles in beyond the SM scenarios to not have any 2-body or 3-body decay channels such that the dominant decay mode is a direct 4-body decay, but from a purely model-independent point of view this should not be discarded off-hand and a likelihood analysis should be performed as in the above examples.}
{\item While extremely unlikely from a theoretical point of view, there is also a possibility that none of the above topology hypotheses provide a good fit such that a direct 5-body decay topology hypothesis may need to be considered.}
\end{itemize}


\section{Conclusions}
\label{sec:conclusions}

With the LHC already operating close to its design energy, it is not unreasonable to expect that even if new physics is discovered, the signal will not have high statistics. Earlier work~\cite{Agrawal:2013uka} demonstrated that for limited signal statistics, kinematic endpoints are inefficient for mass measurements in cascade decays with three visible particles and one invisible particle, and that a determination of the phase space boundary in its full dimensionality can lead to significant improvement. This conclusion was borne out further with a subsequent study with a more realistic analysis~\cite{Debnath:2016gwz}, using the method of Voronoi tessellations~\cite{Debnath:2015wra,Debnath:2016mwb}  to find the boundary of the signal region in the presence of background. In this paper we explored additional decay topologies, including those with four visible particles and one invisible particle, and we have shown that the enhancement in the density of events near the boundary of the kinematically allowed region not only persists, but is even stronger. We have also demonstrated the improvement in mass measurements that can be obtained with these methods on several benchmark decay topologies, for which polynomial methods are not applicable. We have performed this comparison in a very idealized setup, mainly as a proof of principle; however there is no reason to expect that in a more realistic analysis the results obtained by the methods presented in this paper should degrade more than traditional methods based on kinematic endpoints. As has already been done in the case of 4-body final states~\cite{Debnath:2016gwz}, it should be possible to verify whether our conclusions continue to hold using a more realistic analysis. Finally, we have explored the possibility of determining the underlying decay topology using our methods, and we concluded that at least in principle topology determination is achievable. The construction of a more realistic analysis both for mass measurements and for topology determination will be performed in future work.

\acknowledgments
The authors would like to thank Tao Han and Yuan-Pao Yang for valuable conversations regarding this work.  CK and MK are supported by the National Science Foundation under Grant Number PHY-1620610.
The computing for this project was performed on the Momentum Cluster at Northeastern University.

\begin{appendix}
\section{Endpoint formulas}\label{app:endpoints}
In this appendix, we list the formulae for the endpoints used in the analysis of section~\ref{sec:mass}. Additional details and derivations can be found in \cite{Klimek:2016axq}. We work in the limit of massless visible final state particles (except for the lightest partner) for which simple expressions for the endpoints are available.  Numerical verification shows that including small masses has a negligible effect on the endpoint positions.

\subsection{2+2+3}

\begin{figure}
\centering
\includegraphics[scale=0.6]{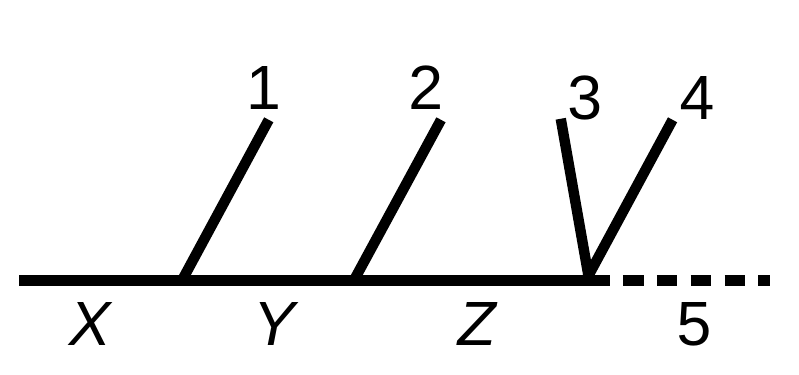}
\caption{\label{fig:223topology}The labeling of final state particles for the 2+2+3 decay topology.}
\end{figure}

The labeling of the particles is illustrated in figure~\ref{fig:223topology}. There are eight endpoints for this topology. The positions of the following four endpoints are spectrum independent: 

\begin{equation}
\max(m_{23}^2) =\max(m_{24}^2) = \frac{(\mys-\mzs)(\mzs-\mfs)}{\mzs},
\end{equation}
\begin{equation}
\max(m_{13}^2) =\max(m_{14}^2) = \frac{(\mxs-\mys)(\mzs-\mfs)}{\mzs},
\end{equation}
\begin{equation}
\max(m_{12}^2) = \frac{(\mxs-\mys)(\mys-\mzs)}{\mys},
\end{equation}
\begin{equation}
\max(m_{34}^2) = (\mz-\mf)^2.
\end{equation}

The positions of the remaining four endpoints are given by expressions that depend on the spectrum: 
\begin{equation}\label{eq:endpoint1234for223}
	\max(m_{1234}^2) = \left\lbrace
	\begin{array}{lcl}
		\frac{(\mxs-\mys)(\mys-\mfs)}{\mys} & \mathrm{if} & \mymf < \mxmy  \\
		\frac{(\mys-\mzs)(\mxs\mzs-\mys\mfs)}{\mys\mzs} & \mathrm{if} & \mymz > \mxmy \mzmf\\
		(\mx-\mf)^2 && \mathrm{otherwise},
	\end{array}\right.
\end{equation}
\begin{equation}\label{eq:endpoint234for223}
	\max(m_{234}^2) =  \left\lbrace
		\begin{array}{lcl}
			\frac{(\mys-\mzs)(\mzs-\mfs)}{\mzs} & \mathrm{if} & \mzmf < \mymz  \\
			(\my-\mf)^2 && \mathrm{otherwise},
		\end{array}\right.
\end{equation}
\begin{equation}
	\max(m_{134}^2) =  \left\lbrace
		\begin{array}{lcl}
			\frac{(\mxs-\mys)(\mzs-\mfs)}{\mzs} & \mathrm{if} & \mzmf < \frac{\sqrt{\mzs+\mxs-\mys}}{\mz}  \\
		 	(\sqrt{\mzs+\mxs-\mys}-\mf)^2 && \mathrm{otherwise},
		\end{array}\right.
\end{equation}
\begin{equation}
	\max(m_{123}^2) =\max(m_{124}^2) = \left\lbrace
		\begin{array}{lcl}
			\frac{(\mxs-\mzs)(\mzs-\mfs)}{\mzs} & \mathrm{if} & \mzmf > \mxmz \\
			\frac{(\mxs-\mys)(\mys-\mfs)}{\mys} & \mathrm{if} & \mymf < \mxmy \\
			\frac{(\mys-\mzs)(\mxs\mzs-\mys\mfs)}{\mys\mzs} & \mathrm{if} & \mymz > \mxmy \mzmf\\
			(\mx-\mf)^2 && \mathrm{otherwise}.
		\end{array}\right.
\end{equation}

\subsection{3+3}

\begin{figure}
\centering
\includegraphics[scale=0.6]{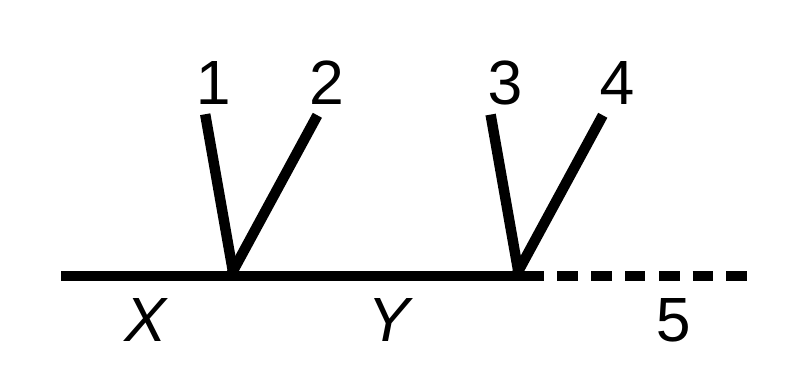}
\caption{\label{fig:33topology}The labeling of final state particles for the 3+3 decay topology.}
\end{figure}

The labeling of the particles is illustrated in figure~\ref{fig:33topology}. There are six endpoints for this topology. The positions of the following four of the endpoints are spectrum independent:

\begin{equation}
	\max(m_{12}^2)=(\mx-\my)^2,
\end{equation}
\begin{equation}
	\max(m_{34}^2)=(\my-\mf)^2,
\end{equation}
\begin{equation}
	\max(m_{13}^2) =\max(m_{23}^2) =\max(m_{14}^2) =\max(m_{24}^2) = (\mxs-\mys)(\mys-\mfs)/\mys,
\end{equation}
\begin{equation}
	\max(m_{1234}^2) = (\mx-\mf)^2.
\end{equation}

The positions of the remaining two endpoints are given by expressions that depend on the spectrum: 
\begin{equation}
	\max(m_{123}^2)=\max(m_{124}^2)=\left\lbrace
		\begin{array}{lcl}
			(\mx-\mf)^2 & \mathrm{if} & \mxmy > \mymf  \\
			\frac{(\mxs-\mys)(\mys-\mfs)}{\mys} & & \mathrm{otherwise}, \\
		\end{array}
	\right.
\end{equation}
\begin{equation}
	\max(m_{134}^2)=\max(m_{234}^2)=\left\lbrace
		\begin{array}{lcl}
			(\mx-\mf)^2 & \mathrm{if} & \mxmy<\mymf  \\
			(\mxs-\mys)(\mys-\mfs)/\mys & & \mathrm{otherwise}.  \\
		\end{array}
	\right.
\end{equation}


\section{Likelihood functions}\label{app:boundarylikelihood}
In this appendix, we will derive analytical expressions for the likelihood functions that we use in our analysis. We treat all particles as spin-0 and we work in the narrow width approximation for any on-shell intermediate states. For a given data sample, we define the likelihood function as the probability that these events were produced from a certain underlying event topology with a spectrum $\{M_{i}\}$ of intermediate on-shell states. Using Bayes' theorem with a flat prior over spectra, one can relate this to the probability of obtaining a given distribution of events from a given spectrum
\begin{equation}
{\mathcal L}(\{M_{i}\}|{\rm data})\propto {\mathcal L}({\rm data}|\{M_{i}\})=\prod_{\rm events} {\mathcal L}_{\rm event}(\{m_{ij}^{2}\}|\{M_{i}\}).
\end{equation}

To capture the multidimensionality of the phase space, we choose ${\mathcal L}_{\rm event}$ factors to be normalized fully differential decay widths, 
\begin{eqnarray}
{\mathcal L}_{\rm event}(\{m_{ij}^{2}\}_{\rm vis.}|\{M_{i}\}) = \frac{1}{\Gamma_X} \int d\Gamma_{X}
\label{eq:likelihooddef}
\end{eqnarray}
integrated over all unobservable $m_{ij}^{2}$ involving the lightest partner\footnote{Note that the visible $m_{ij}^{2}$ are fixed on both sides of equation~\ref{eq:likelihooddef}}. The differential decay width is simply the product of the squared matrix element and the phase space volume element (see equation~\ref{yb_ps_delta}):
\begin{eqnarray}
d \Gamma_X = \frac{|{\mathcal M}|^2}{2 M_X} \, dPS_n \, .
\end{eqnarray}
Since we treat all particles as spin-0, the matrix element squared only contains factors of effective couplings for each decay stage and propagators that are simplified using the narrow width approximation. Therefore, ${\mathcal L}_{\rm event}$ breaks up into factors for each on-shell stage of the cascade decay. Note that each decay stage involves one heavy particle of mass $M_{i}$ that decays to another heavy particle $M_{i+1}$ and a number of light particles, assumed massless.  The energy-momentum conserving $\delta$-functions and factors of $1/\Gamma$ arising from the narrow width approximation for each intermediate on-shell state are also combined with the vertices that they are attached to. See ref.~\cite{Agrawal:2013uka} for additional calculational details.

For 2- and 3-body decay stages, the width of the decaying particle is given by
\begin{eqnarray}
\Gamma_2 &=& \frac{\mu^2}{16 \pi M_i}\LSB 1-r^{2}\RSB \\
\Gamma_3 &=& \frac{\lambda^2 M_i}{512 \pi^3} \LSB 1 - r^4 + 4 r^2 \log (r)  \RSB ,
\end{eqnarray}
where $\mu$ and $\lambda$ are the effective couplings of the 2- and 3-body decay vertices (of mass dimension 1 and 0, respectively), and $r$ is the ratio of the heavy daughter mass to the mass of the decaying particle in that decay stage.

With the correct normalization, the phase space factors for the 4- or 5-body final state in terms of Lorentz invariants are given by
\begin{eqnarray}
dPS_n = M_X^{-2} \left \{
\begin{array}{cc}
2^8 \pi^6 \Delta_4^{-1/2} & n=4 \\
2^{11} \pi^9 \delta(\Delta_5)  & n =5
\end{array}
\right \} \delta(Q^{2}) \prod_{i<j} d(p_{i}\cdot p_{j}) \, ,
\end{eqnarray}
where 
\beq
Q^{2} \equiv \left(\sum_{i<j} p_{i}\cdot p_{j}\right) - \frac{M_X^2-(m_1^2 + \cdots + m_{n-1}^2 + m_{\rm LP}^{2})}{2} = 0
\eeq
encodes overall energy conservation. Here $M_{X}$ is the mass of the decaying particle, $m_{\rm LP}$ is the mass of the lightest partner particle at the end of the decay chain, and the remaining $m_{i}$ are the masses of the light particles in the final state, which we set to zero in our analysis.

Performing the integration over the unobservable $m_{ij}^{2}$, we bring the likelihood functions into a standard form
\beq
\mathcal{L} = \Theta[\mathcal{D}_1] \cdots \Theta[\mathcal{D}_m] \times \mathcal{N} \times \mathcal{F}
\eeq
where for any given decay topology, the $\Theta[\mathcal{D}]$ factors encode the kinematically accessible region, $\mathcal{F}$ contains all dependence on the spectrum $\{M_{i}\}$, and $\mathcal{N}$ includes all remaining dependence on the observed Lorentz invariants $m_{ij}^{2}$ as well as on numerical factors.

In tables~\ref{table:4bodyL} and~\ref{table:5bodyL} we present the likelihood functions for all 4- and 5-body decays consisting of 2- and 3-body decay stages. 
%
We express the results in terms of the kinematic functions $\lambda_n$ with $n(n+1)/2$ arguments, defined as the determinant of a $(n+2)\times(n+2)$ symmetric matrix~\cite{Byckling:1971vca} given by
\begin{eqnarray}\label{eq:definelambda}
\renewcommand{\arraystretch}{1.5}
\lambda_n(x_{1},\ldots,x_{n(n+1)/2}) = \left| \begin{array}{cccccc}
0 & x_1 & x_2 & \cdots & x_{n} & 1 \\
x_1 & 0 & x_{n+1} & x_{n+2} & \cdots & 1 \\
x_2 & x_{n+1} & 0 & x_{2n} & \cdots & 1 \\
\vdots & x_{n+2} & x_{2n} & 0   & \vdots & 1 \\
  & & &   & x_{n(n+1)/2} & 1 \\
x_{n} & & & x_{n(n+1)/2} & 0 & 1 \\
1 & 1 & 1 & \cdots & 1 & 0
\end{array} \right|\, .
\end{eqnarray}
Note that $\lambda_{2}$ is the triangle function that appears in 2-body decays, and $\lambda_n$ is proportional to $\Delta_{n}$ in an $n$-body decay. The likelihood functions can be expressed in terms of $\Delta_i$'s by using the invariant masses $m_{ij}^2$ of pairs of particles, or in terms of $\lambda_n$'s by using the invariant masses of larger collections of particles as shown in tables~\ref{table:4bodyL} and~\ref{table:5bodyL}.  For this reason, the usage of the $\lambda_n$ is superior at making the dependence on the masses of on-shell mediators higher up in the decay chain more explicit, and we adopt this notation in reporting our results.

The structure of the ${\mathcal D}$ factors has interesting properties as well, which we describe in more detail in appendix~\ref{app:factorization}.

\renewcommand{\arraystretch}{1.5}

\begin{table}[!h]
\centering
\begin{tabular}{ | c | c | c | >{$}l<{$} | }
\hline
\multirow{3}{*}{3+2} & \multirow{3}{*}{\raisebox{-20mm}{\includegraphics[width=20mm]{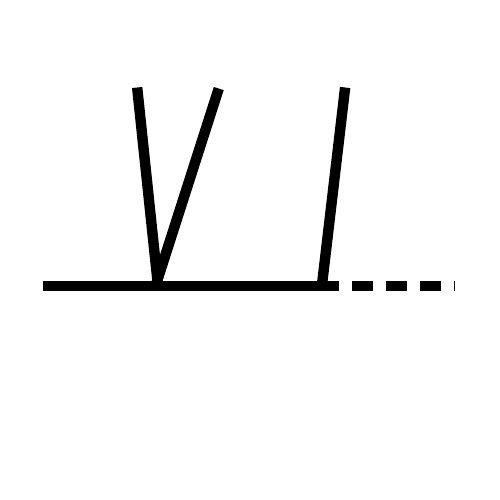}}} & $\mathcal D$ & \lambda_3(m_1^2,m_{12}^2,m_{123}^2,m_{2}^2,m_{23}^2,m_{3}^2) \, \lambda_3(M_X^2,M_Y^2,m_4^2,m_{12}^2,m_{123}^2,m_{3}^2) \\ \cline{3-4}
	& 	& $\mathcal N$ & 16 \, \lambda_2(m_{12}^2,m_{123}^2,m_{3}^2)^{-1/2}  \\ \cline{3-4}
	&	& $\mathcal F$ & M_X^{-4} \, f_3(M_Y/M_X) \, f_2(m_4/M_Y) \\ \hline
\hline
\multirow{3}{*}{2+3} & \multirow{3}{*}{\raisebox{-20mm}{\includegraphics[width=20mm]{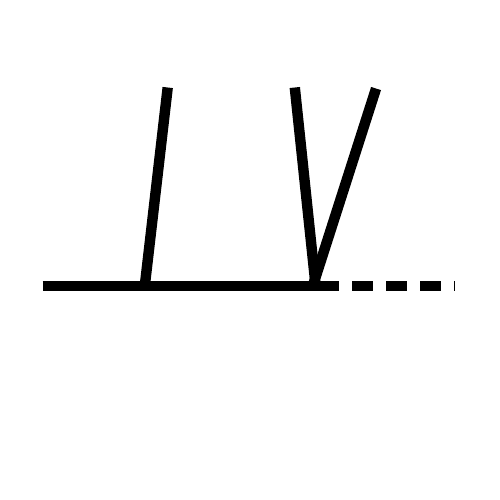}}} & $\mathcal D$ & \lambda_3(m_1^2,m_{12}^2,m_{123}^2,m_{2}^2,m_{23}^2,m_{3}^2) \, \lambda_3(M_X^2,M_Y^2,m_4^2,m_1^2,m_{123}^2,m_{23}^2) \\ \cline{3-4}
	& 	& $\mathcal N$ & 16 \, \lambda_2(m_{1}^2,m_{123}^2,m_{23}^2)^{-1/2} \\ \cline{3-4}
	&	& $\mathcal F$ & M_X^{-2} M_Y^{-2} \, f_2(M_Y/M_X) \, f_3(m_4/M_Y)  \\ \hline
\hline
\multirow{3}{*}{2+2+2} & \multirow{3}{*}{\raisebox{-20mm}{\includegraphics[width=20mm]{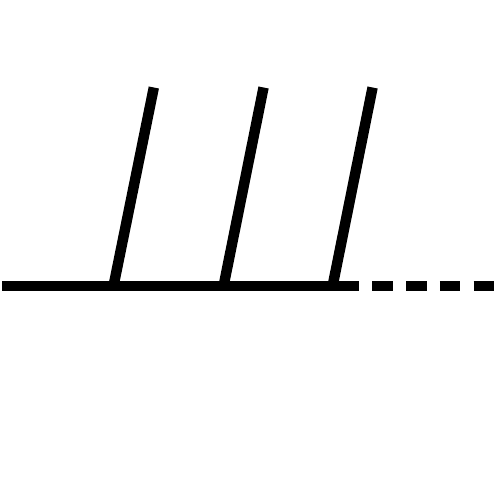}}} & $\mathcal D$ & \lambda_4(M_X^2,M_Y^2,M_Z^2,m_4^2,m_1^2,m_{12}^2,m_{123}^2,m_2^2,m_{23}^2,m_3^2) \\ \cline{3-4}
	& 	& $\mathcal N$ & {8}{\pi^{-1}}   \\ \cline{3-4}
	&	& $\mathcal F$ & M_X^{-2}  \, f_2(M_Y/M_X) \, f_2(M_Z/M_Y) \, f_2(m_4/M_Z) \\&
&& \quad \quad \times \quad \lambda_4(M_X^2,M_Y^2,M_Z^2,m_4^2,m_1^2,m_{12}^2,m_{123}^2,m_2^2,m_{23}^2,m_3^2)^{-1/2}  \\ \hline
\end{tabular}
\caption{\label{table:4bodyL}Likelihood functions for 4-body decays. $f_{2}(r)$ is defined as $(1-r^{2})^{-1}$ and $f_{3}(r)$ is defined as $(1-r^{4}+4r^{2}\log(r))^{-1}$. The $\lambda_{n}$ functions are defined in equation~\ref{eq:definelambda}.}
\end{table}

\begin{table}[!h]
\centering
\begin{tabular}{ | c | c | c | >{$}l<{$} | }
\hline
\multirow{3}{*}{3+3} & \multirow{3}{*}{\raisebox{-25mm}{\includegraphics[width=20mm]{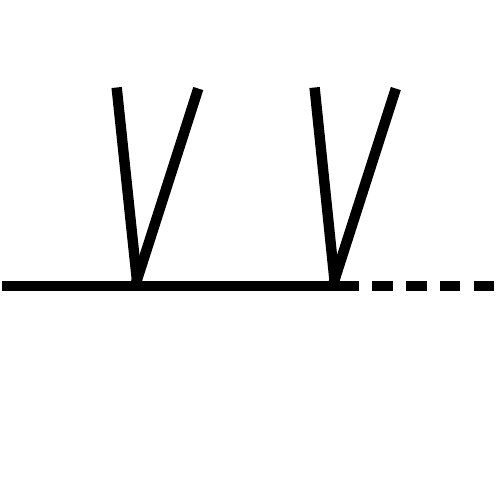}}} & $\mathcal D$ & \lambda_3(m_{12}^2,m_{123}^2,m_{1234}^2,m_{3}^2,m_{34}^2,m_4^2) \, \lambda_3(M_X^2,M_Y^2,m_5^2,m_{12}^2,m_{1234}^2,m_{34}^2)\\ \cline{4-4}
&&& \lambda_4(m_{1}^2,m_{12}^2,m_{123}^2,m_{1234}^2,m_{2}^2,m_{23}^2,m_{234}^2,m_3^2,m_{34}^2,m_4^2) \\ \cline{3-4}
	& 	& $\mathcal N$ &  256\pi^{-1}\, [ \lambda_2(m_{12}^2,m_{1234}^2,m_{34}^2) \\&&& \qquad \times\lambda_4(m_{1}^2,m_{12}^2,m_{123}^2,m_{1234}^2,m_{2}^2,m_{23}^2,m_{234}^2,m_3^2,m_{34}^2,m_4^2) ]^{-1/2} \\ \cline{3-4}
	&	& $\mathcal F$ & M_X^{-4} M_Y^{-2} \,  f_3(M_Y/M_X) \, f_3(m_5/M_Y) \\ \hline
\hline
\multirow{3}{*}{3+2+2} & \multirow{3}{*}{\raisebox{-25mm}{\includegraphics[width=20mm]{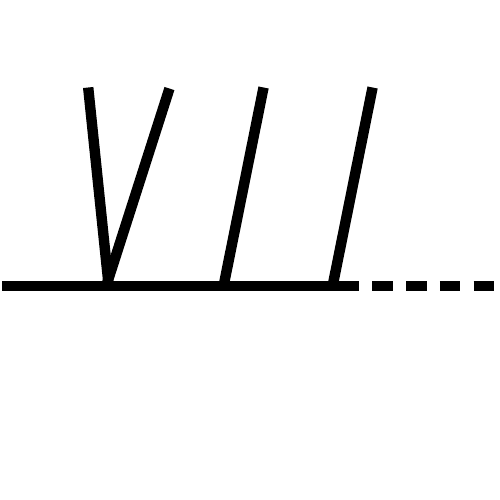}}} & $\mathcal D$ & \lambda_4(m_1^2,m_{12}^2,m_{123}^2,m_{1234}^2,m_2^2,m_{23}^2,m_{234}^2,m_3^2,m_{34}^2,m_4^2) \times  \\&
&& \qquad\lambda_4(M_X^2,M_Y^2,M_Z^2,m_5^2,m_{12}^2,m_{123}^2,m_{1234}^2,m_{3}^2,m_{34}^2,m_{4}^2) \\ \cline{3-4}
	& 	& $\mathcal N$ &  128\pi^{-2} \,  \lambda_4(m_1^2,m_{12}^2,m_{123}^2,m_{1234}^2,m_2^2,m_{23}^2,m_{234}^2,m_3^2,m_{34}^2,m_4^2)^{-1/2} \\ \cline{3-4}
	&	& $\mathcal F$ & M_X^{-4} \, f_3(M_Y/M_X) \, f_2(M_Z/M_Y) \, f_2(m_5/M_Z)  \\&
&& \quad \quad \times \, \lambda_4(M_X^2,M_Y^2,M_Z^2,m_5^2,m_{12}^2,m_{123}^2,m_{1234}^2,m_{3}^2,m_{34}^2,m_{4}^2))^{-1/2}  \\ \hline
\hline
\multirow{3}{*}{2+3+2} & \multirow{3}{*}{\raisebox{-25mm}{\includegraphics[width=20mm]{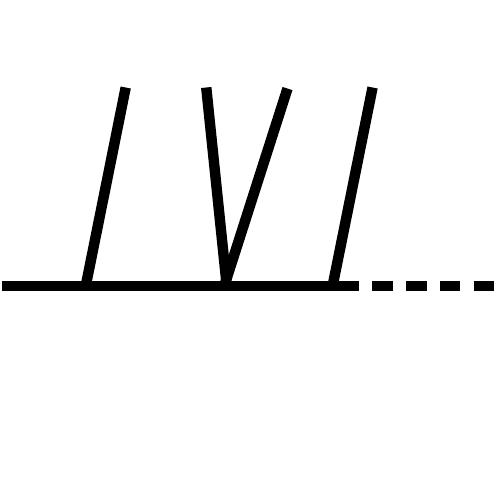}}} & $\mathcal D$ & \lambda_4(m_1^2,m_{12}^2,m_{123}^2,m_{1234}^2,m_2^2,m_{23}^2,m_{234}^2,m_3^2,m_{34}^2,m_4^2) \times  \\&
&& \qquad\lambda_4(M_X^2,M_Y^2,M_Z^2,m_5^2,m_1^2,m_{123}^2,m_{1234}^2,m_{23}^2,m_{234}^2,m_{4}^2) \\ \cline{3-4}
	& 	& $\mathcal N$ & 128\pi^{-2} \,  \lambda_4(m_1^2,m_{12}^2,m_{123}^2,m_{1234}^2,m_2^2,m_{23}^2,m_{234}^2,m_3^2,m_{34}^2,m_4^2)^{-1/2} \\ \cline{3-4}
	&	& $\mathcal F$ & M_X^{-2} M_Y^{-2} \, f_2(M_Y/M_X) \, f_3(M_Z/M_Y) \, f_2(m_5/M_Z)  \\&
&& \quad \quad \times \, \lambda_4(M_X^2,M_Y^2,M_Z^2,m_5^2,m_1^2,m_{123}^2,m_{1234}^2,m_{23}^2,m_{234}^2,m_{4}^2))^{-1/2} \\ \hline
\hline
\multirow{3}{*}{2+2+3} & \multirow{3}{*}{\raisebox{-25mm}{\includegraphics[width=20mm]{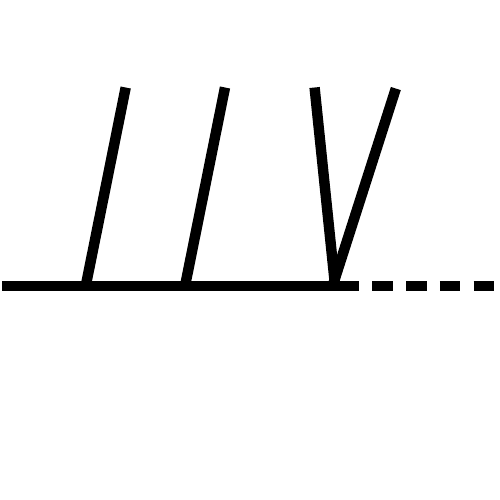}}} & $\mathcal D$ & \lambda_4(m_1^2,m_{12}^2,m_{123}^2,m_{1234}^2,m_2^2,m_{23}^2,m_{234}^2,m_3^2,m_{34}^2,m_4^2) \times  \\&
&& \qquad\lambda_4(M_X^2,M_Y^2,M_Z^2,m_5^2,m_1^2,m_{12}^2,m_{1234}^2,m_2^2,m_{234}^2,m_{34}^2)  \\ \cline{3-4}
	& 	& $\mathcal N$ & 128\pi^{-2} \,  \lambda_4(m_1^2,m_{12}^2,m_{123}^2,m_{1234}^2,m_2^2,m_{23}^2,m_{234}^2,m_3^2,m_{34}^2,m_4^2)^{-1/2}  \\ \cline{3-4}
	&	& $\mathcal F$ &  M_X^{-2} M_Z^{-2} \, f_2(M_Y/M_X) \, f_2(M_Z/M_Y) \, f_3(m_5/M_Z)  \\&
&& \quad \quad \times \, \lambda_4(M_X^2,M_Y^2,M_Z^2,m_5^2,m_1^2,m_{12}^2,m_{1234}^2,m_2^2,m_{234}^2,m_{34}^2))^{-1/2}  \\ \hline
\end{tabular}
\caption{\label{table:5bodyL}Likelihood functions for 5-body decays. $f_{2}(r)$ is defined as $(1-r^{2})^{-1}$ and $f_{3}(r)$ is defined as $(1-r^{4}+4r^{2}\log(r))^{-1}$. The $\lambda_{n}$ functions are defined in equation~\ref{eq:definelambda}.}
\end{table}


\section{Factorization of the domain function}\label{app:factorization}
In this appendix, we will further study the structure of the factors in the likelihood function encoding the kinematically accessible region of phase space. Any cascade decay can be broken down into a number $n_{s}$ of stages, each stage corresponding to the presence of an on-shell intermediate particle. Let us explore how this structure is related to the factorization of the domain function. In particular, consider the $i$-th stage as a heavy particle $X_{i}$ decaying to another heavy particle $X_{i+1}$ and a number $n_{i}$ of SM particles. The domain function cannot depend on whether the $n_{i}$ particles are emitted promptly from the decay vertex, or whether the decay proceeds as $X_{i}\rightarrow X_{i+1} \Sigma_{i}$, with $\Sigma_{i}$ being a metastable particle that much later decays into the $n_{i}$ SM particles\footnote{The mass of the fictitious $\Sigma_{i}$ particle will of course depend on the kinematics of the $n_{i}$ particles in each event.}. In the likelihood function, an essential property of the domain function is to ensure that the full cascade $X_{1}\rightarrow \Sigma_{1}\ldots\Sigma_{n_{s}} X_{n_{s}+1}$ can proceed, where $X_{n_{s}+1}$ is assumed to be the lightest partner particle which is stable.

This consideration gives the key to the factorization of the domain function. There is always a ``skeleton factor'' corresponding to $n_{s}$ consecutive 2-body decays, with the $\Sigma_{i}$ and the lightest partner as final state particles. The skeleton factor cannot be factorized further, and it depends on the spectrum of the $X_{i}$. In addition there are a number of other factors that have to do with the decays of the $\Sigma_{i}$ and these factors depend only on the $m_{ij}^{2}$ of the final state SM particles, but not on the spectrum of the $X_{i}$. Since the $m_{ij}^{2}$ are actually observed in the data, they correspond to a physical configuration of particles and therefore these factors in the domain function can never become negative. In other words, for computing the domain function in the likelihood, only the skeleton factor is nontrivial. The exact form of the remaining factors also depends on the order in which the integrals over the $m_{ij}^{2}$ are performed.

\begin{figure}
\centering
\includegraphics[scale=0.6]{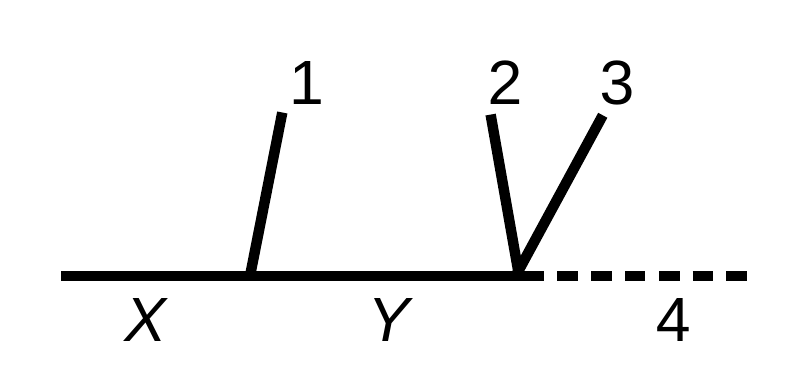}
\caption{\label{fig:23topology}The labeling of final state particles for the 2+3 decay topology.}
\end{figure}

For a concrete example, consider the 2+3 decay topology, where the labeling of the particles is given in figure~\ref{fig:23topology}. $m_{14}^{2}$, $m_{24}^{2}$ and $m_{34}^{2}$ cannot be measured and they need to be integrated over. There are however only two $\delta$-functions arising from on-shell particles $X$ and $Y$ in the narrow width approximation. The phase space volume element also includes a factor of $\Delta_4^{-1/2}$.  Since $\Dfour = -\det M_4$, it is quadratic in all of the $m_{ij}^{2}$. After using the $\delta$-functions to take two of the three integrals, the remaining integral can be performed using the identity
\begin{equation}\label{integral_identity}
	\int_{r_-}^{r_+} \frac{dx}{\sqrt{-ax^2+bx+c}} = \frac{\pi}{\sqrt{a}},
\end{equation}
where $r_\pm$ are the (real) roots of the quadratic expression in the radical. Of course, the identity eq.~\ref{integral_identity} only holds if there exist real roots $r_\pm$, which is equivalent to $\Delta_{4}\ge 0$. This explains why the argument of the domain function is the discriminant of the quadratic expression.

If the last integral is chosen to be over $m_{34}^{2}$, then the discriminant can be factored into two factors $D_{A}$ and $D_{B}$, where $D_{A}$ is the determinant of the $3\times 3$ matrix, the entries of which are dot products of pairs of the four momenta $p_{1}^{\mu}$, $p_{2}^{\mu}$ and $p_{3}^{\mu}$. Similarly, $D_{B}$ is the determinant of the $3\times 3$ matrix, the entries of which are dot products of pairs of the four momenta $p_{1}^{\mu}$, $(p_{2}^{\mu}+p_{3}^{\mu})$ and $p_{Z}^{\mu}$. $D_{B}$ can then be recognized as the skeleton factor, with particles 2 and 3 grouped together into a fictitious $\Sigma$ particle, while $D_{A}$ depends only on the measured $m_{ij}^{2}$. This structure is reflected in the ${\mathcal D}$ entry for the 2+3 topology in table~\ref{table:4bodyL}, with the $\lambda_{3}$ functions corresponding to $D_{A}$ and $D_{B}$.

\end{appendix}

\bibliographystyle{JHEP}

\bibliography{refs}

\end{document}